%% file: course_notes.tex
\documentclass{SciPost}

\hypersetup{
    colorlinks,
    linkcolor={red!50!black},
    citecolor={blue!50!black},
    urlcolor={blue!80!black}
}

\usepackage[bitstream-charter]{mathdesign}
\DeclareSymbolFont{usualmathcal}{OMS}{cmsy}{m}{n}
\DeclareSymbolFontAlphabet{\mathcal}{usualmathcal}

\fancypagestyle{SPstyle}{
\fancyhf{}
\lhead{\colorbox{scipostblue}{\bf \color{white} ~SciPost Physics Lecture Notes }}
\rhead{{\bf \color{scipostdeepblue} ~Submission }}

\fancyfoot[C]{\textbf{\thepage}}
}

\usepackage[english]{babel}
\usepackage{amsmath,bm}
\usepackage{fixmath}

\usepackage{graphicx}
\usepackage{amsthm,graphicx,mathtools,tikz,hyperref}
\usepackage{pgf}
\usepackage{algorithm}
\usepackage{algpseudocode}

\usetikzlibrary{positioning,arrows,intersections}

\graphicspath{{./images/}}

\begin{document}

\pagenumbering{gobble}

\begin{center}{\Large \textbf{\color{scipostdeepblue}{
Numerical Cosmology
}}}\end{center}

\begin{center}\textbf{
Romain Teyssier
}\end{center}

\begin{center}
Department for Astrophysical Sciences\\
4 Ivy Lane, Princeton University\\
Princeton, 08544, New Jersey\\
United States of America
\\[\baselineskip]
\href{mailto:teyssier@princeton.edu}{\small teyssier@princeton.edu}
\end{center}

\section*{\color{scipostdeepblue}{Abstract}}
\boldmath\textbf{%
In these lecture notes, we describe the current state-of-the-art for numerical simulations of large-scale structure and galaxy formation.
Numerical simulations play a central role in the preparation and the exploitation of large-scale galaxy surveys, in which galaxies are the fundamental observational objects. We first describe basic methods for collisionless N-body dynamics that enable us to model dark matter accurately by solving the Vlasov-Poisson equations. We then discuss simple methods to populate dark matter halos with galaxies, such as Halo and Sub-halo Abundance Matching techniques and baryonification techniques for capturing baryonic effects on the matter distribution. We finally describe how to model the gas component by solving the Euler-Poisson equations, focusing on the foundational assumptions behind these equations, namely local thermodynamical equilibrium, and the nature of the truncation errors of the numerical scheme, namely numerical diffusion. We show a few examples of simulations of a Milky-Way-like halo without cooling, with cooling and with star formation. We finally describe different subgrid prescriptions recently developed to model star formation, supernovae feedback and active galactic nuclei and how they impact cosmological simulations.
}

\vspace{\baselineskip}

\noindent\textcolor{white!90!black}{%
\fbox{\parbox{0.975\linewidth}{%
\textcolor{white!40!black}{\begin{tabular}{lr}%
  \begin{minipage}{0.6\textwidth}%
    {\small Copyright attribution to authors. \newline
    This work is a submission to SciPost Physics Lecture Notes. \newline
    License information to appear upon publication. \newline
    Publication information to appear upon publication.}
  \end{minipage} & \begin{minipage}{0.4\textwidth}
    {\small Received Date \newline Accepted Date \newline Published Date}%
  \end{minipage}
\end{tabular}}
}}
}


\newcommand{\dd}{\mathrm{d}}
\newcommand{\pdd}[2]{\frac{\partial #1}{\partial #2}}
\newcommand{\odd}[2]{\frac{\dd #1}{\dd #2}} 
\newcommand{\ldd}[2]{\frac{D #1}{D #2}}
\newcommand{\bvec}[1]{\mathbf{#1}}
\newcommand{\tensor}[1]{\mathbb{#1}}

\vspace{10pt}
\noindent\rule{\textwidth}{1pt}
\tableofcontents
\noindent\rule{\textwidth}{1pt}
\vspace{10pt}

\include{introduction}
\include{chapter1}

\include{chapter2}

\include{chapter3}

\include{chapter4}

\bibliography{romain.bib}

\end{document}

%% file: introduction.tex
\section{Introduction}

The purpose of this course is to give an introduction to the concepts and techniques used in the field of {\it Numerical Cosmology} to model the matter distribution in the Universe on large scales. This is particularly timely with the advent of multiple experiments aiming at mapping the distribution of galaxies to unprecedented accuracy. The Euclid satellite has been flying for a couple of years and is acquiring data for its first data release, while the Vera Rubin telescope is seeing its first photons and collecting giant images of the sky at the exact same time as our Summer school. In a slightly more distant future, the Roman space telescope will hopefully fly and contribute to this deluge of data that many call the Golden Age of cosmology.

Given the shear volume of data, the term {\it precision cosmology}, usually reserved to observations of the  Cosmic Microwave Background (CMB) , can be now extended to large scale galaxy surveys. It turns out that precision cosmology is much more challenging in the realm of galaxy surveys than it is for the CMB. The key difference sits in the regime of the gravitational dynamics: matter fluctuations at the epoch of the last scattering are in the linear regime of the gravitational instability, while they are deep into the non-linear regime at the epoch of the galaxies we see today around us. This puts numerical methods at the center of the game and the associated numerical errors as an important source of systematic errors.

In these lecture notes, we will unfortunately only graze the surface of the huge amount of material one has to cover to become an expert in numerical cosmology. We suggest the interested reader to read additional material in these famous textbooks or follow the trail of references mentioned later in the manuscript.
 
\begin{itemize}
\item The Large-Scale Structure of the Universe by P. J. E. Peebles
\item Galaxy Formation and Evolution by H. Mo, F. van den Bosch, and S. D. White
\item Modern Cosmology by S. Dodelson
\item Galaxy Formation by M. S. Longair
\item Cosmologie: Des Fondements Théoriques aux Observation by J. P. Uzan (in French)
\end{itemize}

This course will be divided into 4 parts, each associated with one lecture during the Summer school:
\begin{enumerate}
\item Collisionless N body Dynamics
\item From Dark Matter Haloes to Galaxies
\item Baryonic Physics and Hydrodynamics
\item Stars, Black Holes and Feedback
\end{enumerate}

The first part covers the foundation of dynamics of dark matter, which we believe is mostly collisionless. We will detail the master equations describing its evolution across cosmic ages, together with simple approximations such as Eulerian and Lagrangian perturbation theories and the spherical collapse model. The second part will describe numerical techniques used to detect automatically the building block of the Universe in the non-linear regime of the gravitational instability, namely dark matter haloes. These halo finders are fundamental tools to turn these complex simulations into mock galaxy catalogs used to prepare and interpret observations. These catalogs can be "observed" by a virtual telescope using the "light cone" technique. Once we have these halo catalogs, we can populate them with galaxies using simple "galaxy painting" techniques that turns out to be surprisingly accurate. We will cover in particular abundance matching, regulator modelling and semi-analytical modeling. These techniques are useful for quick simulations of the galaxy population but cannot replace direct hydrodynamical simulations of the baryonic (gaseous) component. In the third part, we will describe in details the equations governing the dynamics of baryons, namely the Euler equations, as well as key physical processes such as radiative cooling and heating. These ingredients allow in principle the self-consistent formation of galactic disks from first principles, which is a very exciting and very ambitious goal of numerical cosmology. Unfortunately, we now know better. Our last chapter will cover the formation of stars and black holes and the associated numerical methods. These processes are obviously not resolved by our cosmological simulations. This is why we refer to these ingredients as "subgrid models". It turns out that modeling the formation of stars and black holes and in particular their associated energy and momentum feedback mechanism is fundamental to cosmology on large scales. This feedback loop couples sub-parsec scales to mega-parsec scales and affect directly the cosmological signal we would like to extract from these galaxy surveys. This is the realm of baryonic effects that act as a very serious systematic effects on the large scale distribution of matter in the Universe. Finally, our conclusion will cover very briefly some missing aspects such as non-standard model of gravity and dark matter and their associated numerical techniques.

%% file: chapter1.tex
\section{Collisionless N-body Dynamics}

\subsection{Introduction}

In this first chapter, we describe numerical techniques used to model the dynamics of dark matter in the expanding Universe.  Dark matter particles are modeled as a collisionless fluid, following the so-called Vlasov-Poisson equations \cite{Efstathiou.1985, Angulo.2022}. The theoretical background leading to the proper justification of the validity of the Vlasov-Poisson approach is highly non-trivial and has been covered elsewhere \cite{Rampf.2021}. In summary, in order to have a proper collisionless system of N particles, with N very large, one has to fulfill two main conditions: 1- The gravitational acceleration on each particle has to be dominated by the long-range, mean or fluid gravitational field contribution, and the short-range, two-body contribution should be negligible, and 2- The initial particle distribution should have no spatial correlations, meaning the 2-points probability distribution function (PDF) is just the product of both one-point PDFs. The second condition is less known and can prove problematic for N body codes \cite{Binney.2002,Diemand.2004}. 

We can write the Vlasov-Poisson system as
\begin{equation}  
\frac{\partial f}{\partial t} + {\bm v} \cdot \frac{\partial f}{\partial {\bm x}} + {\bf g} \cdot \frac{\partial f}{\partial {\bm v}} = 0
\end{equation}
where $f$ is the distribution function, that gives the number of particles per phase space volume element:
\begin{equation}
{\rm d}N = f\left({\bf x}, {\bf v}, t\right) {\rm d}^3 x{\rm d}^3 v
\end{equation}
and ${\bf g}$ is the {\it mean field} gravitational interaction provided by the solution of Poisson's equation
\begin{equation}
\Delta \phi = 4\pi G \rho~~{\rm and}~~~{\bf g} = - \nabla \phi
\end{equation}
The right-hand side of Poisson's equation features the gravitational constant $G$ and the fluid dark matter mass density defined as the first velocity moment of the distribution function as
\begin{equation}
\rho \left( {\bf x}, t \right) = m_{\rm dm} \int f\left({\bf x}, {\bf v}, t\right) {\rm d}^3 v
\end{equation}
We unfortunately don't know the dark matter particle mass $m_{\rm dm}$ or any other of its fundamental properties \cite{Nadler.2019} but we can absorb this constant in the definition of the distribution function substituting  $f \rightarrow m_{\rm dm} f$. Even though we failed to detect it in any laboratory experiment so far \cite{Aprile.2016, Misiaszek.2024}, the most popular dark matter particle candidate is the WIMP as Weakly Interacting Massive Particle. Its large mass allows us to assume its initial distribution in phase space is an infinitesimally thin sheet with zero ``temperature'', meaning in this case the phase-space distribution is a delta function in velocity space. These cold dark matter model, together with a few key parameters describing the expansion rate of the Universe across cosmic ages, are the core ingredients of the popular Lambda Cold Dark Matter model or LCDM for short.

\subsection{Vlasov-Poisson in Phase-Space}

The most natural numerical technique for solving the Vlasov-Poisson equations is to discretize phase space into volume elements defined as
\begin{equation}
 V_{i, j}= [x_{i-1/2}, x_{i+1/2}]\times [v_{j-1/2},v_{j+1/2}]
\end{equation}
with a constant mesh size $\Delta x = x_{i+1/2}-x_{i-1/2}$ and $\Delta v = v_{j+1/2}-v_{j-1/2}$.
The time evolution of the distribution function can be discretized using a conservative formulation as:
\begin{equation}
f_{i,j}^{n+1} = f_{i,j}^n -\frac{\Delta t }{\Delta x} ( F_{i+1/2,j}-F_{i-1/2,j} ) -  \frac{\Delta t}{\Delta v}(G_{i,j+1/2}-G_{i,j-1/2})
\end{equation}
where time is also discretized using $t^{n+1}=t^n + \Delta t$. The intercell fluxes can be computed using a simple first-order upwind scheme as:
\begin{equation}
F_{i+1/2,j}=v_j f_{i,j}~~~{\rm if}~~~v_j>0~~~{\rm or}~~~F_{i+1/2,j}=v_j f_{i+1,j}~~~{\rm if}~~~v_j<0
\end{equation}
and
\begin{equation}
G_{i,j+1/2}=g_i f_{i,j}~~~{\rm if}~~~g_i>0~~~{\rm or}~~~G_{i,j+1/2}=g_i f_{i,j+1}~~~{\rm if}~~~g_i<0
\end{equation}
Note that in the previous equations, we have used the fact that the acceleration depends only on the position ${\bf g}({\bf x}, t)$ which is why our N body system is symplectic, or in other words its Hamiltonian is separable. We still need to compute the acceleration at each time step using any of the field solvers we discuss later in this Chapter. 

The upwinding for the flux functions is a crucial aspect of {\it advection schemes}. This ensures that the numerical solution will remain positive and the time integration stable. Indeed, it is easy to show that for constant $v$ and $g$ with $v>0$ and $g>0$, the previous numerical update can be written as:
\begin{equation}
f_{i,j}^{n+1} = f_{i,j}^n \left( 1 - \frac{\Delta t }{\Delta x} v - \frac{\Delta t }{\Delta v} g \right)  + f_{i-1,j}^n  \frac{\Delta t }{\Delta x} v +  f_{i,j-1}^n \frac{\Delta t}{\Delta v}g 
\end{equation}
which is a convex combination of the solution at time $n$, provided the time step satisfies the Courant stability condition:
\begin{equation}
\Delta t \left (\frac{v}{\Delta x} + \frac{g}{\Delta v} \right)  < 1
\end{equation}
This simple scheme is only first-order accurate. It suffers from numerical errors (also called numerical diffusion) depending on the phase-space resolution $(\Delta x, \Delta v)$ but also on the phase-space boundaries via $v_{\rm max}$ and $g_{\rm max}$. Higher order methods featuring polynomial reconstruction of the distribution function at the cell edges can greatly improve the quality of the solution \cite{Sousbie.2016,Tanaka.2017}. In the context of LCDM, however, initial conditions are singular and the razor thin sheet in velocity space is a challenge for classical phase-space solvers.

Another challenge for phase-space solvers is the sheer dimensionality of the problem. If we aim at $N_x=N_v=32$ resolution elements per dimension, which is clearly not glorious, this translates into $N=N_x^3 N_v^3$ elements in phase space or 1 billion cells. The overall computational cost scales as ${\rm max}(N_x, N_v){\cal O}(N)$, owing to the Courant stability condition. The most spectacular example of Vlasov simulations can be found in \cite{Supinski.2021}. 

A new interesting technique has been developed recently by \cite{Hahn.2013} trying to follow the dynamics of this thin manifold directly in phase-space. The resulting code works beautifully at early times when the sheet is folding over itself a small number of times \cite{Ondaro-Mallea.2023}. As time goes by, unfortunately, the number of sheet folding increases exponentially, rendering the method intractable.

\subsection{Vlasov-Poisson using Particles}

Even if recent advances in supercomputing power would allow one to implement the previous phase-space solvers, particle-based methods have proven superior and easier to implement in practice \cite{Hockney.1988}. Particle methods sample phase-space using point masses defined by their coordinates in phase space at time $t^n$:
\begin{equation}
\left( {\bf x}_p^n, {\bf v}_p^n\right)_{p=1, N}
\end{equation}
where now $N$ stands for the total number of particles. Particle positions and velocities are updated using the {\it equation of motion}:
\begin{equation}
\frac{d{\bf x}_p}{dt} = {\bf v}_p~~~{\rm and}~~~ \frac{d{\bf v}_p}{dt} = {\bf g}_p
\end{equation}
where the time derivative is now the Lagrangian time derivative, which is the time derivative of the corresponding quantity {\it while moving with the particle}, as opposed to the Eulerian time derivative, which is the time derivative of the corresponding quantity {\it at a fixed point in phase-space}. The acceleration ${\bf g}_p$ is also the gravitational acceleration {\it at the position of the particle}. In particle methods, the gravitational acceleration can be computed using various field solvers described below. 

Note that these particles have nothing in common with actual dark matter particles. They are discrete macroscopic mass element sampling an initial region of phase-space and subsequently moving as a single giant point mass. This discrete Lagrangian view is pretty far from the correct fluid description of dark matter and various discreteness effects are cause for concern. 

On the positive side, however, particles allow a straightforward description of a razor-thin sheet in phase-space. We only need to populate a lattice of $N=N_x^3$ particles with say $N_x=32$, which translate in having {\it only} 32'768 resolution elements, compared to 1 billion for the phase-space solver. The velocity of each particle can be set in the initial condition, effectively defining the geometry of the razor-sheet phase-space manifold. 

\subsection{Field Solvers}

Solving for Poisson's equation requires different strategies whether one uses a particle-based or a cell-based approach. In the former, the gravitational acceleration felt by a single particle $p$ can be obtained straightforwardly using a {\it direct summation} method as:
\begin{equation}
{\bf g}_p({\bf x}_p) = - G m \sum_{q \neq p} \frac{{\bf x}_p - {\bf x}_{q}}{\left| {\bf x}_p - {\bf x}_{q}\right|^3}
\end{equation}
where $m$ is the mass of the macro-particles. The simplicity and the accuracy are its two main advantages. But this direct summation approach suffers from two major issues. First, its costs is not sustainable for very large values of $N$. Indeed, the algorithmic complexity scales as ${\cal O}(N^2)$, since for each particle $p$, we have a loop over all other particles $q$. Second, if 2 particles come arbitrarily close to each other, the resulting binary interaction will dominate over the mean field acceleration. We will therefore violate the first condition for the system to be collisionless and ruin the validity of the Vlasov-Poisson equation. The second problem can be fixed using a softened gravity defined as:
\begin{equation}
{\bf g}_p({\bf x}_p) = - G m \sum_{q \neq p} \frac{{\bf x}_p - {\bf x}_{q}}{\left| {\bf x}_p - {\bf x}_{q}\right|^3+\epsilon^3}
\end{equation}
where $\epsilon$ is the so-called {\it softening parameter}. In computational cosmology, it is traditionally set to $\epsilon = \Delta x/50$ where $\Delta x$ is the spacing of the initial particle lattice. This number is obviously a free parameter that needs to be adjusted depending on the exact nature of the initial conditions and on the number of particles.

If one wants to speed up direct summation of the gravitational interactions, the method of choice is the Tree Code (Barnes-Hut algorithm \cite{Barnes.1986}). The idea of the tree code is to decompose the computational space in a recursive octree. Each cell of the octree is subdivided in 8 new cells (hence the name octree) until it contains a small enough number of particles, say less than 32. Starting from the leaf cells upward towards the tree trunk, a single cell covering the entire computational domain, the octree accumulates in each cell its mass, center of mass and even higher order multipoles. 

The multipole expansion is used during the acceleration calculation by grouping distant particles into their parent tree cells, according to a tree cell {\it opening criterion}. If a tree cell is too close to the particle whose acceleration is being computed, the tree cell is opened and the process is repeated with the children tree cells. Nearby cells are opened until individual particle are exposed and their contribution to the acceleration is identical to the direct summation. Distant cells, however, contribute to the acceleration via a multipole expansion of the potential, therefore speeding up drastically the computation compared to the direct summation method.  The algorithmic complexity of the Tree Code is ${\cal O}(N \ln N)$, the exact number of operations depending on the opening criterion. 

The Fast Multipole Method variant (Greengard-Rokhlin algorithm \cite{Greengard.1987}) is even more efficient as it does not only consider interactions between octree cells and particles but also cell-cell interactions. It also makes use of a multipole expansion of the acceleration, not just the mass distribution. The algorithmic complexity  is theoretically  ${\cal O}(N)$, but again the exact number of calculations depends on the details of the mass distribution. 

For the phase-space solver, the natural approach is to solve directly Poisson's equation on the same grid as the one we used to define the mass density. We have $N=N_x^3$ cells of size $\Delta x$ for which the density is $\rho_i$ where each cell is labeled by index $i$. We then approximate the Laplace operator using a Finite Difference (FD) approximation which reads (shown here only in 1D):
\begin{equation}
\frac{\phi_{i+1}-2\phi_i+\phi_{i-1}}{\Delta x^2} = 4\pi G \rho_i
\end{equation}
We therefore need to solve a linear system of equation that requires ${\cal O}(N^2)$ operations if one uses the traditional Gauss elimination strategy, which is obviously very inefficient. The Conjugate Gradient method works better in this case (since we have a symmetric positive definite matrix) and gives us a faster time-to-solution with an algorithmic complexity of ${\cal O}(N)$ per iteration. The Conjugate Gradient is indeed an iterative method, the problem being here that the number of iteration depends on the condition number of the matrix, which for Poisson's equation depends on the grid resolution. So the number of iterations required to reach a certain convergence criterion increases with the grid resolution, resulting in an increased cost of  ${\cal O}(N^{4/3})$.  

The real game changer in this context is the Fast Fourier Transform. This algorithm, invented by Cooley and Tukey \cite{Cooley.1965} in the mid-sixties, allows one to perform the Fourier transform on a discrete grid in ${\cal O}(N \ln N)$ operations. It is based on an odd-even recursive division of Fourier modes. This technique is particularly suited to our FD approximation of the Laplace operator, since its Fourier transform becomes a diagonal matrix, trivial to invert:
\begin{equation}
-\frac{4}{\Delta x^2}\sin^2 \left(\frac{k\Delta x}{2} \right) {\hat \phi}_k = 4 \pi G {\hat \rho}_k
\end{equation}
The gravitational acceleration is obtained similarly using a FD approximation of the gradient operator as:
 \begin{equation}
g_i = - \frac{\phi_{i+1}-\phi_{i-1}}{2\Delta x}
\end{equation}
For very large grid size, however, the $N \ln N$ scaling can become prohibitive. A faster algorithm has been proposed by Brandt called Multigrid \cite{Brandt.1977}. Multigrid is another iterative method exploiting a hierarchy of grid of different resolution (similar to the octree described above) with interpolation and averaging operators allowing the solution to converge quickly at different scales using a so-called Jacobi smoother. The Multigrid method scales as ${\cal O}(N)$ and outperforms the FFT for large $N$.

\subsection{The Particle-Mesh Method}

Given the speed of the FFT or the Multigrid method, it is tempting to use those field solvers for particle-based methods too. This is the main motivation at the origin of the Particle Mesh (PM) method \cite{Hockney.1988}, for which the following sequence of operators is applied:
\begin{enumerate}
\item Deposit the particle masses onto the mesh and compute the grid-based density field
\item Solve for Poisson's equation on the mesh (FFT, FMM or Multigrid)
\item Compute the acceleration on the mesh
\item Interpolate the grid-based acceleration at the particle positions
\end{enumerate}
The two critical steps of the PM scheme are 1- mass deposition and 4- force interpolation. For this, we need to provide a way to represent the density and acceleration field as continuous fields, so that they are defined everywhere in space, not just at the discrete set of mesh points. The traditional method for accurate interpolation is based on {\it spline interpolants}. The density and acceleration at any point in space $x$ are given by:
 \begin{equation}
\rho(x) = \sum_{i=0}^p \rho_i B^p(x-x_i)~~~{\rm and}~~~g(x) = \sum_{i=0}^p g_i B^p(x-x_i)
\end{equation}
where the sum is over the $p+1$ basis elements of the piecewise spline polynomials of degree $p$. The spline coefficients $\rho_i$ are obtained by projecting the particle distribution onto the spline basis using the standard inner product for polynomials, namely
\begin{equation}
\rho_i = \sum_{p=1}^N \int B^p(x-x_i)\delta(x-x_p) {\rm d}x  
\end{equation}
and $\delta$ is the Dirac distribution centered on each particle. Standard mass deposition schemes are Cloud-In-Cell ($p=1$), Triangular-Shaped-Cloud ($p=2$) and Piecewise-Cubic-Spline ($p=3$). The higher the polynomial degree, the more basis elements and therefore mesh points are involved in the interpolation.  The higher cost comes usually with the benefits of a less noisy density field and a more isotropic gravitational acceleration. 

Many hybrid methods have been developed combining these different techniques, such as the Particle-Mesh/Particle-Particle method \cite{Davis.1985} (P$^3$M) or the Tree Code \cite{Hernquist.1987} or the Tree/Particle-Mesh method \cite{Bode.2000} (TPM) or the Adaptive Particle-Mesh method \cite{Bryan.1997,Bryan.1997} (APM), combining fast field solvers (usually based on FMM or Multigrid) and high force accuracy (using adaptive mesh refinement or small force softening).

\begin{figure}[httb]
\centering
\includegraphics[scale=0.28]{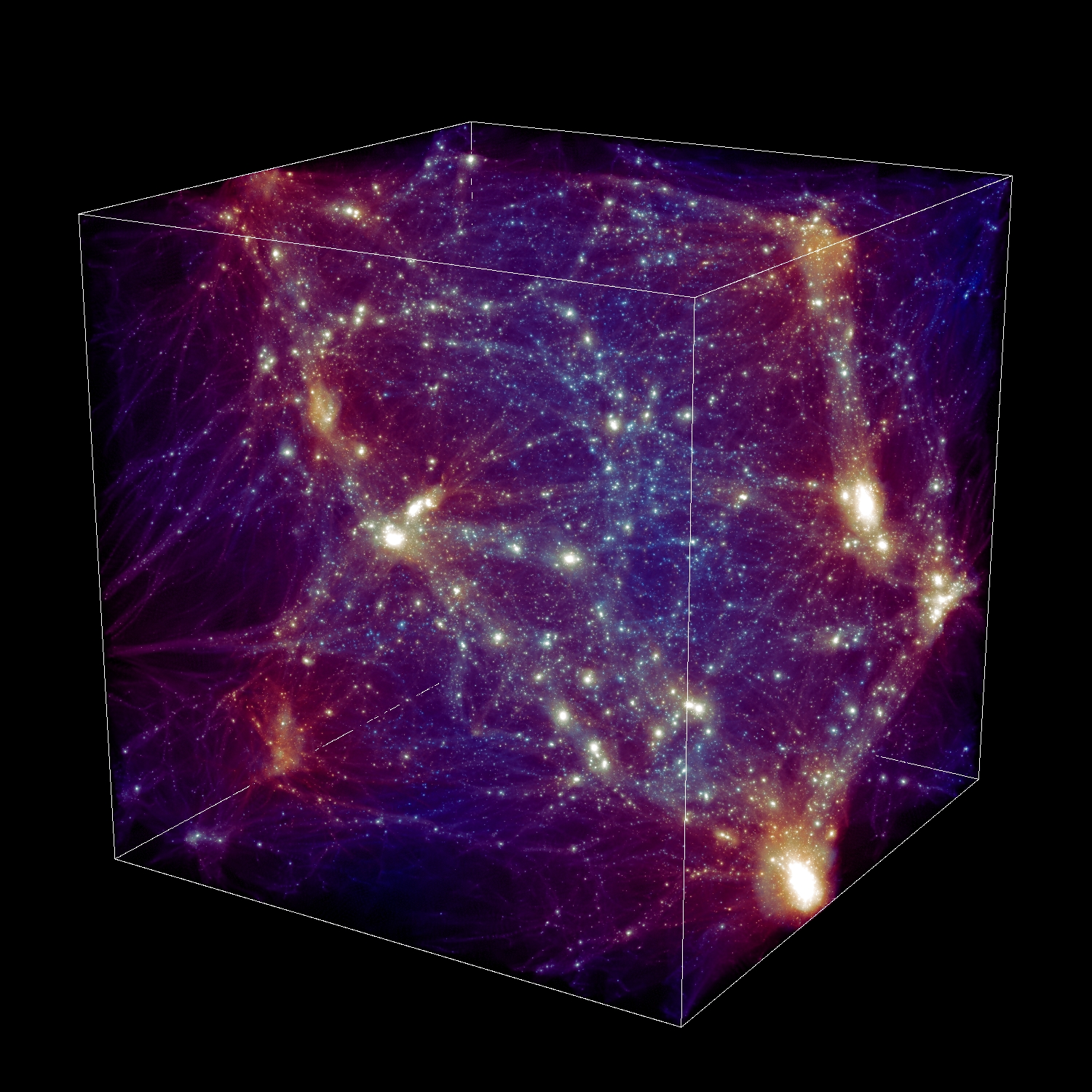}
\caption{Color rendering of the dark matter particle distribution from a N-body simulation of a LCDM model. Dark matter halos are visible as clumps of different sizes. The color coding follows the mass of the halos, with large halos appearing redder and small halos appearing bluer. The characteristic filamentary structure of the Cosmic Web is striking. Simulation credit: RAMSES. Image credit: S. Colombi.}
\label{fig:colombi}
\end{figure}

\subsection{Symplectic Time Integrators}

Once we know the gravitational acceleration at the position of each particle, we need to integrate the equation of motion. This is not such an obvious task, as we would like to preserve one key property of symplectic N body systems, namely the strict conservation of energy. For sake of simplicity, we will consider here the harmonic oscillator, the archetypical symplectic system. The equation of motion write in this case:
\begin{equation}
\frac{dv}{dt}=-\omega^2 x~~~{\rm and}~~~\frac{dx}{dt} = v
\end{equation}
The simplest time integration is given by the Forward Euler scheme by:
\begin{equation}
v^{n+1} = v^n - \omega^2 x^n \Delta t~~~{\rm and}~~~x^{n+1} = x^n + v^n \Delta t 
\end{equation}
We can try and compute the total energy at time $t^{n+1}=t^n + \Delta t$ as:
\begin{equation}
E^{n+1} = \frac{1}{2}(v^{n+1})^2 + \frac{1}{2}\omega^2 (x^{n+1})^2 = E^n \left( 1 + \omega^2 \Delta t^2\right) 
\end{equation}
We see that the energy of each particle is increasing by the factor $1+\omega^2 \Delta t^2$ every time step, which is an undesirable numerical effect. The consequence will be that particle orbits will never remain bound and our collisionless system will forever expand.  A very simple fix can be found using a minor modification of the previous scheme called Symplectic Euler:
\begin{equation}
v^{n+1} = v^n - \omega^2 x^n \Delta t ~~~{\rm and}~~~x^{n+1} = x^n + v^{n+1} \Delta t 
\end{equation}
We see that instead of a single step where we update both the velocity and the position together, we now update first the velocity alone, performing the so-called {\it kick step}, followed by the {\it drift step}, during which we update the position alone, using the new velocity instead of the old one. 
We now define a modified total energy, also know as the {\it shadow Hamiltonian} as:
\begin{equation}
E^{\prime} = \frac{1}{2}v^2 + \frac{1}{2}\omega^2 x^2 + \frac{1}{2}\omega^2 x v \Delta t
\end{equation}
It is easy to show that Symplectic Euler satisfies (left to the reader as an exercise):
\begin{equation}
(E^{\prime})^{n+1} = (E^{\prime})^n
\end{equation}
Such a time integrator is called symplectic \cite{Wisdom.1991} because it conserves exactly a first-order accurate approximation of the true total energy. By combining the proper sequence of kick and drift steps, one can design higher order symplectic schemes, such as the famous Leap Frog or Verlet Scheme \cite{Stoermer.1907,Verlet.1967}:
\begin{eqnarray}
v_{1} & = & v^n - \omega^2 x^n \Delta t/2\\
x^{n+1} & = & x^n + v_{1} \Delta t\\
v^{n+1} & = & v_{1} - \omega^2 x^{n+1} \Delta t/2
\end{eqnarray}
which features an intermediate velocity state noted here $v_1$ at the half time step $t^n+\Delta t/2$. One can show that the Leap Frog scheme conserves exactly the following second-order shadow Hamiltonian (left to the reader as a more difficult exercise):
\begin{equation}
E^{\prime} = \frac{1}{2}v^2 + \frac{1}{2}\omega^2 x^2 - \frac{1}{8}\omega^4 x^2 \Delta t^2
\end{equation}
Another scheme is Yoshida's Fourth-Order Scheme \cite{Yoshida.1990} which writes
\begin{eqnarray}
v_1 & = & v^n - \omega^2 x^n \frac{\lambda}{2} \Delta t\\
x_1 & = & x^n +v_1 \lambda \Delta t\\
v_2 & = & v_1 - \omega^2 x_1 \frac{1-\lambda}{2} \Delta t\\
x_2 & = & x_1 +v_2 (1-2\lambda) \Delta t\\
v_3 & = & v_2 - \omega^2 x_2 \frac{1-\lambda}{2} \Delta t\\
x^{n+1} & = & x_2 + v_3 \lambda \Delta t\\
v^{n+1} & = & v_3 - \omega^2 x_3 \frac{\lambda}{2}\Delta t
\end{eqnarray}
where the constant $\lambda \simeq 1.351207$ is the real root of the polynomial $-1+6\lambda-12\lambda^2+6\lambda^3=0$. The interested and highly motivated reader can try and find the corresponding shadow Hamiltonian, using the Baker–Campbell–Hausdorff expansion. Note however that the intermediate states use negative time steps, an annoying property shared by all symplectic integrators of order strictly higher than 2.

These results have been derived for the simple case of the harmonic oscillator, but they can be generalized to any collisionless N body system, as long as the acceleration depends only on the particle positions and the corresponding dynamics is symplectic. An interesting consequence of the strict conservation of the modified energy is the unconditional stability of symplectic schemes. The choice of a small enough time step is only motivated by accuracy, not by stability. This is a major difference with the phase-space upwind scheme we derived earlier, where stability is the decisive factor controlling the size of the time step. The attentive reader may have noticed that the 3 symplectic schemes we have presented in this section are all {\it time reversible}. Indeed, if we perform a new time integration starting with the final particle positions and velocities and reverting the time step from $\Delta t$ to $-\Delta t$, we  obtain exactly the initial particle positions and velocities. This is another specific property of symplectic time integrators.

\subsection{High Performance Computing}

The different techniques we have presented in this Chapter have been implemented in multiple codes over the last 4 decades. The most recent implementations of the tree code, especially the FMM variant, are available in the {\ttfamily GADGET} \cite{Springel.2021} and {\ttfamily PKDGRAV} \cite{Potter.2017} codes. These codes have different versions, mostly a massively parallel implementation using the Message Passing Library (MPI) or OpenMP compiler directives, and a Graphical Processing Unit (GPU) implementation on modern architecture such as NVidia chips. The most recent implementations of the P3M code are available in the {\ttfamily HACC} \cite{Habib.2016} and {\ttfamily ABACUS} \cite{Garrison.2021} codes. They also take advantage of the most recent GPU architectures. The most recent implementations of the Adaptive PM code are available in the {\ttfamily RAMSES} \cite{Teyssier.2002}, {\ttfamily ENZO} \cite{Bryan.2014} and {\ttfamily ART} \cite{Kravtsov.1997} codes. Note that only a fraction of these N-body codes have also coupled hydrodynamics solvers, as discussed in the third Chapter. Most of these codes are also Open Source and can be freely downloaded from GitHub or BitBucket repositories. 

Cosmological simulations require very large supercomputers. Two types of simulations are usually necessary to prepare and extract the scientific exploitation of large galaxy surveys. The first type, called grand challenge runs, features one (or a few) gigantic simulation that provides both good mass and spatial resolution and very large simulated volumes. Upcoming surveys aim at resolving galaxies down to 1/10 of the mass of the Milky Way, requiring a particle mass resolution of $m \simeq 10^9$~M$_\odot$ (with a minimum of 100 particle per halo). A box size of 200~Mpc/h corresponds to a simulation of $1024^3$ or 1 billion particles. This is too small of volume for any type of reasonable cosmological application. For example, the Euclid Flagship simulation performed with the {\ttfamily PKDGRAV} code has a box size of 3.6~Gpc/h and $16000^3$ or 4 trillion particles. The second type, called simulation suites, performed a large number of smaller runs. For example, the AbacusSummit simulation suite produced almost 140 simulations with $6000^3$ particles and 2 Gpc/h box sizes, using the {\ttfamily ABACUS} code. 

Historically, it is interesting to analyze the growth of the number of particles simulated as a function of the publication year. It is a clear exponential growth with doubling the number of particles every year. This is very fast, faster than Moore's law describing the growth of the number of floating point operations per seconds on the best supercomputers available that same year. The reason is now well known: Each decade has seen the advent of a new computer architecture (scalar computers in the 80s, vector computers in the 90s, massively parallel computer's in the 2000s and now GPU computers in 2010s until now) but also of a new N-body algorithm (PP (direct N-body) in the 70s, PM and P3M in the 80s, Tree code in the 90s, Adaptive PM in the 2000s, FMM and Multigrid in the 2010s), with scaling improving steadily from ${\cal O}(N^2)$ to ${\cal O}(N \ln N)$ and finally to ${\cal O}(N)$.

\subsection{Systematic Effects and Numerical Errors}

Now that we understand better the challenges of the stability, accuracy and computational cost of different N-body methods, we will discuss briefly possible systematic errors to have in mind when trying to compute accurate theoretical predictions. We have very few rare example of analytical predictions that can be used to test numerical codes. The best method so far, deep into the non-linear regime where analytical predictions do not exist, is to compare the predictions of different codes using different techniques. As illustrated by \cite{Schneider.2016} and \cite{Garrison.2018}, existing high-resolution methods seems to agree with the linear regime on large scale and seems to agree with each other on small scale to better than 1\% up to a wavenumber $k \simeq 10$~h/Mpc. At smaller scales, details in the time integrators or in the field solvers start to differ enough so the inherently chaotic dynamics make the different solutions diverge. It is still unclear if N-body codes could be made to agree on such small scales.

Another source of systematic errors is that fact that most (if not all) N-body code are non-relativistic. A notable exception is the {\ttfamily GEVOLUTION} code \cite{Adamek.2016} that can deal properly with all relativistic corrections of large scale structure formation. The danger here is to create an artificial difference between the linear predictions of relativistic, multi-fluid Boltzmann solvers (such as CAMB \cite{Lewis.2011} or CLASS \cite{Blas.2011}) and the predictions of N-body code on large scales, where linear theory is valid. In order to correct for this spurious effect, one has to rely on techniques such as {\it backscaling} or even include as an additional smooth background component the relativistic perturbations (photons, neutrinos and metric) directly into the N-body solver (see \cite{Fidler.2018,Tram.2018,Brandbyge.2018}).

Finally, at very small scales, close to the initial particle lattice Nyqvist frequency $k_{\rm Nyq} \simeq \frac{\pi}{\Delta x}$, particle discreteness effects produce spurious results in key observables such as the power spectrum or the bi-spectrum. A pragmatic approach could be to just ignore all modes beyond say half of the Nyqvist frequency, literally ignoring modes that required so much computing power to obtain. Another approach is to use {\it particle linear theory}, dealing explicitly with spurious modes introduced by the crystal-like structure of the initial particle lattice, and correct them in the power spectrum \cite{Joyce.2009}. This very sophisticated approach, combined with additional tricks detailed in other lectures of the book, allows one to exploit N-body data all the way down to the Nyqvist frequency \cite{Garrison.2016}.

\subsection{Beyond N-body}

Finally, we will discuss an important line of products produced by modern N-body simulations, namely alternative models to LCDM. Modified gravity theories such as MOND \cite{Milgrom.1983} or $f(R)$ gravity \cite{Capozziello.2011}, rely on a non-linear version of Poisson's equation \cite{Li.2012,Puchwein.2013,Lueghausen.2015}. Future large-scale galaxy surveys have in principle the power to detect any small deviation from Newtonian gravity embedded in the standard Einstein theory of gravity. Being able to interpret these potential deviations requires to have codes producing accurate predictions within the framework of these alternative models of gravity. The main technical difficulty here is that non-linear modifications of Poisson's equation prevent us from using traditional Green's function convolution approaches such as FFT, tree code or FMM. The only method that we can use in this context is the non-linear version of the Multigrid method. 

Other alternative models of LCDM explore the possibility for dark matter to be self-interacting (SIDM) \cite{Spergel.2000} or made of very light bosons (Fuzzy Dark Matter, FDM) \cite{Hu.2000}. These models require different levels of modification of the basic N-body method outlined in this Chapter. SIDM models need to implement some kind of collision tracking, remove momentum or energy during each event \cite{Hu.2000}. FDM models need to solve the Schroedinger-Poisson equations \cite{Schive.2014}, which share a surprisingly large number of properties with Vlasov-Poisson, with however a key difference in the numerical requirement to resolve spatially and temporally the dynamics of the wave function. These methods give interesting predictions at small scales, offering interesting alternative to LCDM in case future observations reveal possible deviations in the dark matter dynamics.

%% file: chapter2.tex
\section{Halos Finding and Mock Galaxy Catalogs}

\subsection{Introduction}

N-body simulations provide an invaluable insight into the non-linear dynamics of dark matter. On large scale, matter is organized in sheets and filaments interconnected in a so-called Cosmic Web. While cosmic voids are the most salient features in term of volume, most of the mass in the Universe accumulates at the nodes of the filaments in dense compact clumps called ``halos''. Although cosmic voids, filaments and walls are also non-linear objects whose dynamics and detailed structures are still the topics of intense research \cite{Kraljic.2017}, the evolution of the gravitational instability deep into its non-linear regime is traditionally described using a halo-based model, or in short the ``Halo Model'' \cite{Cooray.2002}. Another strong motivation for adopting this discrete, halo-centric view is the connection to the galaxy population. Observational cosmology (also known as extragalactic astronomy) is indeed based mostly on the statistical analysis of the properties and spatial distribution of distant galaxies. Constraining the cosmological model using past, on-going and future large galaxy surveys requires to predict the position and luminosity of the entire galaxy population. Our current view is that each dark matter halo hosts a central galaxy at its center, and several smaller satellite galaxies in its outer regions. Predicting the exact positions and properties of galaxies within their host halos would require to understand the process of galaxy formation. This is an entirely different ball game that we will discuss in the next chapter. Intermediate solutions, such as ``abundance matching''  \cite{Wechsler.2018} or ``semi-analytical models'' \cite{Somerville.2015} allow us to produce very realistic mock galaxy catalogs without the need for full physics modeling of galaxy formation. Combined with N-body simulations, this simplified, halo-based approach of galaxy formation, allows to produce very accurate theoretical predictions for galaxy clustering and weak lensing.

\subsection{Spherical Collapse and the Virial Theorem}

An exact and commonly accepted definition of dark matter halos has surprisingly not emerged yet. There is a rough agreement on the quantitative properties of dark mater halos, but these properties can vary depending on the context. The formation of dark matter halos can be described theoretically using several powerful analytical tools: 1- thshe spherical collapse model \cite{Gunn.1972},  2- the Press \& Schechter (PS) statistical theory \cite{Press.1974}, 3- the Navarro, Frenk \& White (NFW) density profile \cite{Navarro.1997}.  These analytical models are unfortunately only approximate and need to be calibrated using N-body simulations. We can also adopt a more pragmatic approach and use various numerical algorithms to define halos directly in these non-linear simulations. We will discuss these different aspects here, and use this discussion as a motivation for a proposal of a strict definition of dark matter halos that can be hopefully adopted by the community once and for all.

The spherical collapse model considers an initial spherical region of constant overdensity $\delta_i$ and comoving radius $R$. This region will collapse at (or before) time $t$ if its initial overdensity {\it linearly extrapolated} at time $t$ satisfies:
\begin{equation} 
\delta = \frac{D^+(t)}{D^+(t_i)} \delta_i > \delta_c=\frac{3}{5}\left( \frac{3\pi}{2} \right)^{2/3} \simeq 1.686
\end{equation}
where $D+(t)$ is the linear growth rate of the initial perturbation in the adopted cosmology. Note that the previous formula is valid only for an Einstein-de Sitter Universe ($\Omega_m=1$). The value is $\delta_c \simeq 1.675$ for the LCDM model.
The final halo mass is the Lagrangian mass of the initial region:
\begin{equation}
M = \frac{4\pi}{3}\rho_m R^3
\end{equation}
where $\rho_m = \Omega_m \rho_c$ is the average mass density in the Universe and $\rho_c$ is the critical density. The spherical collapse model only predict the trajectory of spherical shells until the collapse, defined as the time where the density becomes infinite. In order to estimate the density of the halo {\it after the collapse}, the spherical collapse model uses a rather fallacious argument based on the following story. In reality, the collapse is not spherical and the density is not strictly uniform. The collapse leads to a complex process called ``violent relaxation'' during which the dark matter particles are redistributed in phase-space into an equilibrium configuration where the velocity dispersion of the particles balances the gravity of the halo. This thermal equilibrium is described by the virial theorem, that states that an isolated system in equilibrium must satisfy $2K+U=0$ where $K$ is the particle kinetic energy and $U$ is the halo gravitational potential energy. Although the spherical collapse model cannot describe this final relaxed state, it however briefly cross the point $2K+U=0$ when the uniform density of the collapsing sphere is:
\begin{equation}
\rho_{\rm vir} = 18\pi^2 \rho_c \simeq 178 \rho_c = 178 \rho_m
\end{equation}
This last formula being valid only for an Einstein-de Sitter Universe for which $\rho_m=\rho_c$. For LCDM, we have $\Omega_m \simeq 0.3$ and the virial overdensity is (see e.g. \cite{Bryan.1998}):
\begin{equation}
\rho_{\rm vir} \simeq 101 \rho_c = 337 \rho_m
\end{equation}
Numerical simulations tell us a different story. Dark matter halos do not form out of the monolithic collapse of a spherical homogeneous sphere. Interestingly, dark matter halos appear to be indeed in relatively good virial equilibrium in their inner regions. Virial equilibrium is a condition for equilibrium of a spherical isolated system between the kinetic energy and the potential energy.
\begin{equation}
2E_{\rm kin} + E_{\rm grav} = 0~~~{\rm or}~~~V_{\rm circ}^2 \simeq \frac{G M}{R_{\rm vir}}
\end{equation}
where $V_{\rm circ}$ is the average rms velocity of dark matter particles. If these particles are on circular orbits, this is their orbital velocity. Hence, we call this velocity the circular velocity $V_{\rm circ}$. Under virial equilibrium, the dark matter density remains roughly constant and the mean radial velocity vanishes. In the outskirts, however, this is not the case and one can measure non-zero radial infalling flows. The transition between the equilibrium, virialized inner regions and the infalling outer region seems to be defined in simulations using a different overdensity criterion, the value depending on the halo mass and the epoch \cite{Cuesta.2008}. The non-linear dynamics within the central virialized region is in the so-called ``stable clustering'' regime with a constant dark matter physical density and a zero-radial velocity fully decoupled from the expansion of the Universe.

\subsection{Halo Statistics}

\begin{figure}
\centering
\includegraphics[scale=0.3]{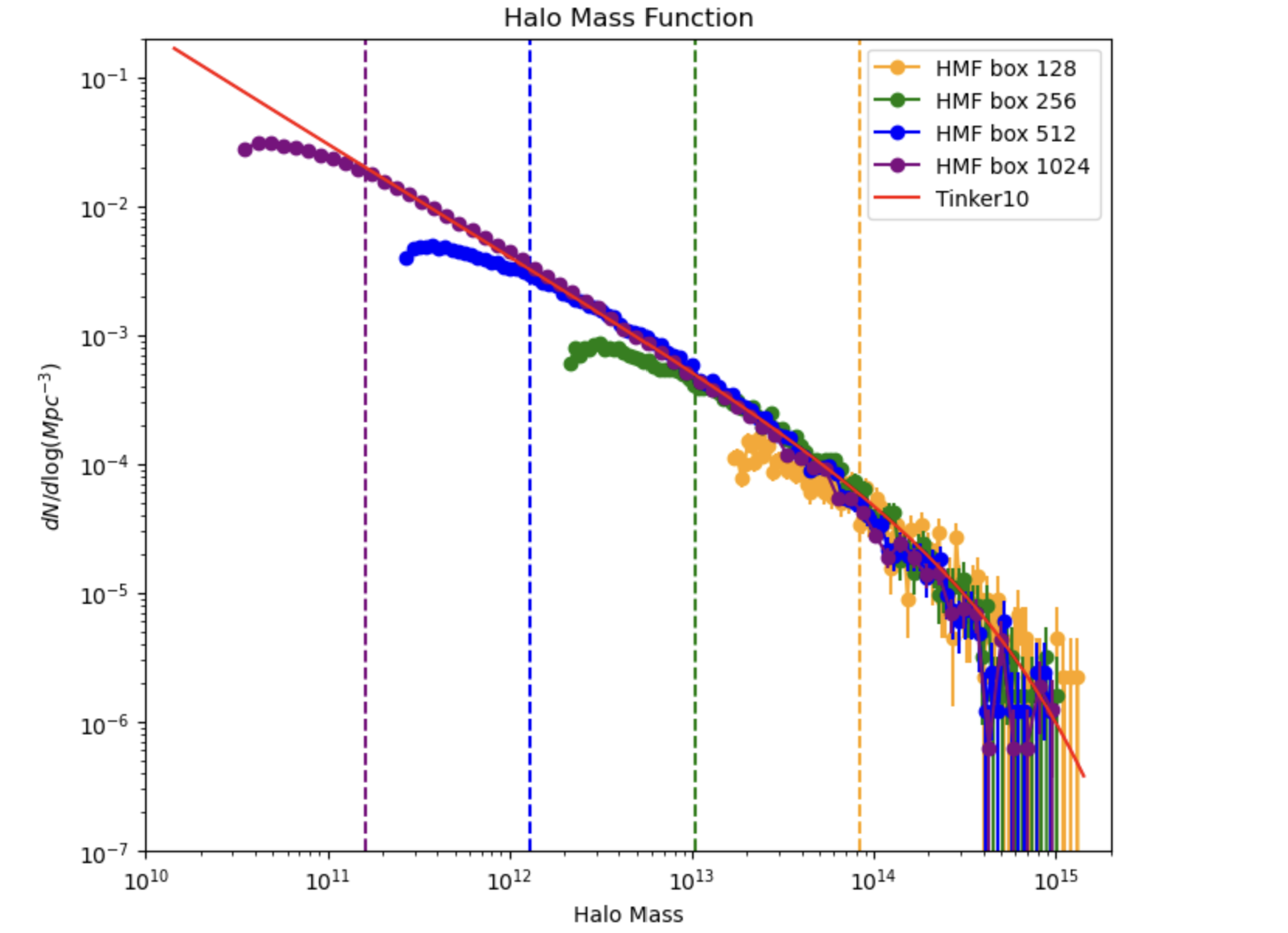}
\caption{Halo mass function from 200~Mpc/h N-body simulations with different particle number (from $128^3$ to $1024^3$). We used Poisson error bars in each mass bin. The vertical dashed lines indicate the 100 particles per halo limit below which we cannot trust the simulations. The solid red line is the analytical fit from Tinker (2010). Credit figure: R. Ait Ekioui. Credit simulation: R. Teyssier.}
\label{fig:numhmf}
\end{figure}

The second strategy to define dark matter halos is via their statistical properties. The previously discussed spherical regions form from very specific initial conditions, namely Gaussian random fields. In this context, Press \& Schechter (PS) theory \cite{Press.1974} identifies the mass fraction in the Universe locked inside collapsed halos of mass larger than $M$ as {\it twice} the volume fraction of spherical regions with linearly extrapolated overdensity larger than the spherical collapse threshold:
\begin{equation}
F(>M) = 2 \int_{\delta_c}^{+\infty} \frac{1}{\sqrt{2\pi}} \exp\left( -\frac{\delta^2}{2\sigma^2(M)}\right)\frac{1}{ \sigma(M)}{\rm d}\delta
\end{equation}
where $\sigma(M)$ is the variance of the density field {\it smoothed} at the comoving scale $R$ corresponding to the mass of the halos. The factor of 2 is a dirty trick so that the mass fraction is properly normalized to 1 as $M \rightarrow 0$. The resulting halo mass function is obtained via:
\begin{equation}
\frac{dn}{dM} = -\frac{\rho_m}{M}\frac{dF}{dM} = - \sqrt{\frac{2}{\pi}} \, \frac{\rho_m}{M} \, \frac{\delta_c}{\sigma^2} \frac{d \sigma}{d M}  \exp\left( -\frac{\delta_c^2}{2 \sigma^2} \right)
\label{eq:pstheory}
\end{equation}
and gives the number density of halos of mass between $M$ and $M+{\rm d}M$. $\sigma(M)$ is obtained using the power spectrum of the initial Gaussian random fluctuations via:
\begin{equation}
\sigma^2(M) = \frac{1}{8\pi^3} \int_0^{+\infty}4\pi k^2P(k)W(kR){\rm d}k
\end{equation}
where $W(kR)$ is the Fourier transform of the Top-Hat smoothing filter. This simple formula turns out to be remarkably accurate when compared to numerical simulations. We show in Figure~\ref{fig:numhmf} the halo mass function (HMF) in a 200 Mpc/h periodic box simulated with different particle numbers (from $128^3$ to $1024^3$) compared to the Tinker analytical fit \cite{Tinker.2008}. The underlying theory appears a bit shaky at first. It is now much more complete after several decades of improvement, via the so-called Extended-Press-Schechter (EPS) theory \cite{Bower.1991,Lacey.1993}. This leads us to consider now a very important question: how do we define halos in N-body simulations?

The first attempt to detect automatically halos in the mid-80s was motivated by the spherical collapse model, in particular the spherically average overdensity $\Delta_{\rm vir} =\rho_{\rm vir}/\rho_m = 178$. The first algorithm to detect halos was introduced at that time \cite{Davis.1985,Jenkins.2001}: Friend-Of-Friends (FOF). The idea is to collect particles into groups defined as regions of space with density larger than a given threshold. Mathematically, halo number $i$ is defined by the volume: ${\bf x}\in V_i$ for which $\rho({\bf x})> \rho_{\rm min}$, bonded by the isodensity surface: ${\bf x}\in S_i$ for which $\rho({\bf x})= \rho_{\rm min}$. The FOF algorithm proceeds as follows:
\begin{enumerate} 
\item Choose a linking length $h$
\item Start with one particle with coordinate ${\bf x}_p$
\item Add to a list all particles for which $\left| {\bf x}_q-{\bf x}_p\right| < h$
\item Repeat with all particles in the list until there is no more particle to add
\item Repeat for the next group
\end{enumerate}
It is easy to show that the corresponding overdensity density is:
\begin{equation}
\rho_{\rm min} = \frac{6 m}{4\pi h^3}
\end{equation}
since particles of mass $m$ are added to a group if we can find at least 2 particles within a sphere of radius $h$. If one uses a linking length $h=0.2\Delta x$ where $\Delta x$ is the initial particle grid resolution, also known as the {\it mean inter-particule spacing}, we find $\rho_{\rm min} \simeq 60 \rho_m$, since by definition $m = \rho_m \Delta x^3$.

\begin{figure}
\centering
\includegraphics[scale=0.4]{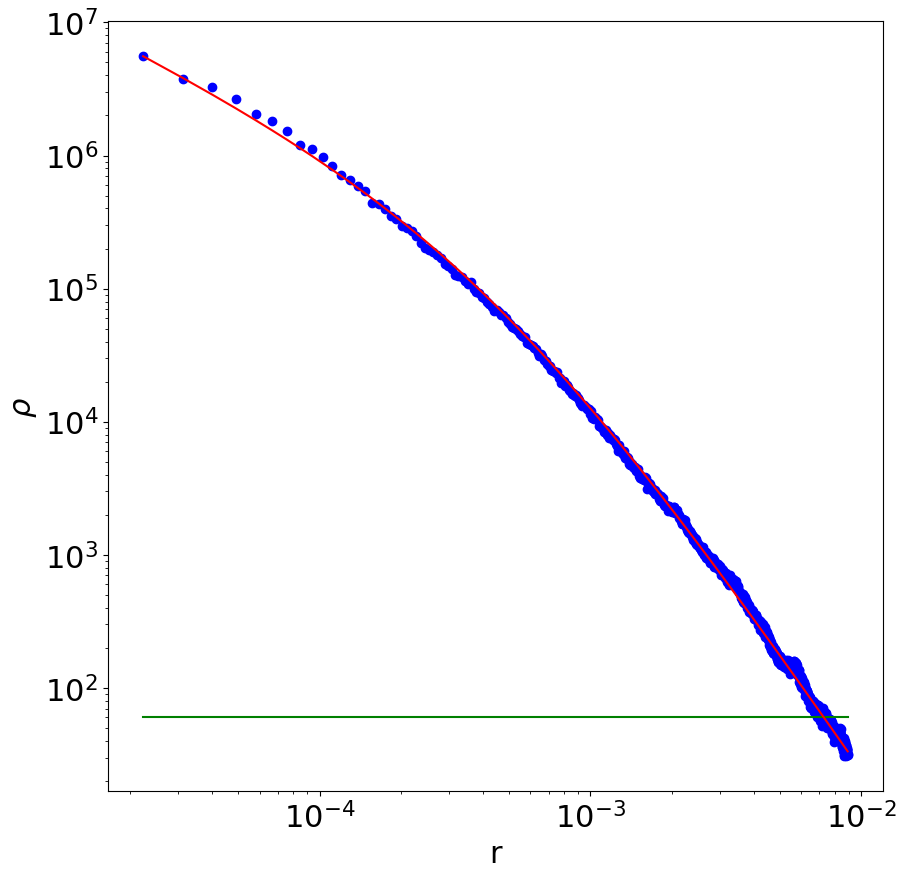}
\caption{Density profile of the simulated halo shown in Figure~\ref{fig:dmo_zoom} (blue circles) compared to the analytical NFW profile with $c=20.5$. The isodensity level $\rho_{\rm min}=60$ is shown as the horizontal green line. The innermost blue point corresponds to twice the resolution of the grid. Credit figure and simulation: R. Teyssier.}
\label{fig:nfw}
\end{figure}

How do we get from $\rho_{\rm min} \simeq 60\rho_m$ to $\Delta_{\rm vir} \simeq 180 \rho_m$? This is where things get a little shaky. We assume the typical halo profile follows the Singular Isothermal Sphere (SIS) model with:
\begin{equation}
\rho(r) = \rho_s \left( \frac{r}{r_s}\right)^{-2},
\end{equation}
so that we can compute the cumulative mass within radius $r$ as:
\begin{equation}
M_{\rm tot}(r) = 4\pi \rho_s \rho_s r_s^2 r.
\end{equation}
The average overdensity within radius $r$ is finally:
\begin{equation}
\Delta(r) = \frac{3 M_{\rm tot}(r)}{4\pi r^3} = 3 \rho(r)
\end{equation}
We thus find that to match $\Delta_{\rm vir} \simeq 180 \rho_m$, we have to choose an isodensity contour $\rho_{\rm min} \simeq 60 \rho_m$, and therefore a linking length of $h \simeq 0.2 \Delta x$. We know however from N-body simulations that the SIS model is not an accurate description of the density profile of dark matter halo. The halo density profile is in fact described to great accuracy by the NFW profile:
\begin{equation}
\rho(r) =  \frac{\rho_s}{\frac{r}{r_s}\left( 1+ \frac{r}{r_s}\right)^2}.
\end{equation}
The slope is -1 in the center and -3 in the outskirts (see Fig.~\ref{fig:nfw}), so overall the SIS model and its -2 slope remains a good approximation to define the linking length and detect halos. If one wants to compute the ratio between the average overdensity and the isodensity contour, one gets:
\begin{equation}
\frac{\Delta(r)}{\rho(r)} = 3 \left( \ln (1+x)-\frac{x}{1+x}\right)\frac{(1+x)^2}{x^2} ~~~{\rm where}~~~x=\frac{r}{r_s}
\end{equation}
The results depends on the value adopted for $\Delta_{\rm vir}$ and on the concentration parameter $c$ of each halo, defined by:
\begin{equation}
c = \frac{R_{\rm vir}}{r_s}.
\end{equation}
For example, $c\simeq 5$, which corresponds to the typical concentration of the largest halos in the Universe, gives something closer to $\Delta_{\rm vir} \simeq 3.8 \rho_{\rm min}$, while $c\simeq 10$, which corresponds to Milky Way analogues, gives $\Delta_{\rm vir} \simeq 5.4 \rho_{\rm min}$. It seems impossible in practice to choose a unique value for $\rho_{\rm min}$ that matches every halo concentration.

Another argument for choosing $\rho_{\rm min} \simeq 60 \rho_m$ and $h \simeq 0.2 \Delta x$ comes from the statistics of dark matter halos. Indeed, N-body simulations have shown that when one adopts $\rho_{\rm min}=60 \rho_m$ for the FOF algorithm linking length parameter, the resulting halo mass function matches the PS theoretical prediction almost magically well \cite{Jenkins.2001}. More importantly, if one varies the cosmology or the cosmic epoch for the simulated universe, the numerical mass function satisfies a key property of the PS prediction, namely the mass function {\it self-similarity}. Indeed, one can see easily in Equation~\ref{eq:pstheory} that the mass function depends only on $\sigma(M)$ which encodes the linear matter power spectrum. If one express the mass function as a function of the self-similar variable $\nu=D^+(t) \delta_c/\sigma(M)$, it becomes a function of only $\nu$, therefore independent on both time and cosmology. This key property can only be obtained if one chooses $\rho_{\rm min} \propto \rho_m$ and the resulting mass function is the closest to PS theory if one chooses precisely $\rho_{\rm min}= 60 \rho_m$.

In order to reconcile these conflicting definitions, the following strategy seems to emerge as the proper way to define halos in N-body simulations. 
\begin{enumerate}
\item Use FOF 0.2 or any grouping scheme based on the density isosurface $\rho_{\rm min} = 60 \rho_m$ to detect halos \cite{Jenkins.2001},  and define their masses as $M=M_{200m}$ using $\Delta_{\rm vir} = 200 \rho_m$, even at different epochs and for different cosmologies
\item Use the virial definition $\Delta_{\rm vir} = 200 \rho_c$ to define their mass as $M=M_{200c}$, where index $200c$ means the halo mass and radius corresponds to $\Delta_{\rm vir} = 200 \rho_c$, even at different epochs and for different cosmologies
\end{enumerate}
This dual definition sounds a bit schizophrenic, but it guarantees both a sound statistics, with a self-similar mass function and a good agreement with PS theory if one use $M=M_{200m}$, and it guarantees a sound equilibrium dark matter particle distribution if one use $M=M_{200c}$. It is of course essential to always define properly what density isosurface value has been used to define the halo mass and radius, hence the use of index $200c$ or $200m$ everywhere deemed appropriate. 

\subsection{Centrals and Satellites}

Defining halos via a density threshold $\rho_{\rm min}$ allows to segment dark matter particles into distinct, non-overlapping regions that can roughly match a typical overdensity $\Delta_{\rm vir}$. This is a very good first step but unfortunately this is not precise enough if one wants to go beyond the halo number density. For example, if one needs to compute the halo density profile, and compare it to the NFW model for example, the obvious first step is to define the halo center. This is where things get ugly. The center of mass of the halo region, for example, is a well define coordinate but it rarely matches the density maximum within the halo region. Computing the density profile accurately requires the center to coincide with at least a local density maximum, if not the global density maximum within the halo region \cite{Gelb.1994,Klypin.1999}. 

Another problem arises during halo mergers. Indeed, two approaching halos will have initially well separated density isosurfaces at $\rho_{\rm min}$, but the two regions will merge into a single region as soon as the two isosurfaces touch each other. Even though the center of the two halos are still separated by the sum of their respective virial radii, the algorithm considers them as a single halo region. If the mass of the two halos is comparable, say within a factor of a few from each other, there is no reason to choose one over the other as the center of the system, and the second as a satellite of the first one within the system. Only when the two peaks will actually merge can we agree that the two halos have merged fully.

The proper approach is to segment the dark matter density field using a proper ``peak patch'' decomposition. Morse theory provides a rigorous mathematical framework for this task by classifying critical points (maxima, minima, and saddles) of a scalar field and constructing the Morse-Smale complex, which partitions space based on gradient flow. This approach underlies tools like \texttt{DisPerSE} \cite{Sousbie.2011}, which extract topologically robust features using persistence homology. In contrast, algorithms like \texttt{Subfind} \cite{Springel.2001}, \texttt{AHF} \cite{Knollmann.2009} and \texttt{AdaptaHOP} \cite{Tweed.2009} are density-based and heuristic, identifying peaks and substructures without explicitly using topological constructs. The Watershed algorithm, commonly used in image processing, offers an intuitive analogy to Morse theory by segmenting fields into basins around local extrema. This principle is employed in void-finders like \texttt{ZOBOV} \cite{Neyrinck.2008} and in the \texttt{PHEW} algorithm within the \texttt{RAMSES} code \cite{Bleuler.2015}. \texttt{PHEW} performs a hierarchical watershed segmentation of the density field on AMR grids, identifying peaks and merging them based on saddle densities. While inspired by Morse theory, \texttt{PHEW} does not compute the full Morse-Smale complex or use persistence, but it provides a scalable and efficient method for structure detection in large N-body simulations. One of the most widely used algorithm, \texttt{ROCKSTAR} \cite{Behroozi.2012}, takes advantage of the phase-space information to identify halos. It is particularly well suited to detect weak disrupted satellites, although it might not align so well with the requirements of large galaxy surveys.

Each halo region is thus segmented hierarchically into several peak patches. When the largest peak patches in the hierarchy are more massive than say 10\% of the total halo mass, they are considered as ``central`` peaks, while less massive peak patches are considered as ``satellite`` peaks. Each peak is surrounded by a well defined ``peak patch`` surrounded by its saddle surface. This region can be used to compute the mass associated to the peak patch and to compute the density profile around the peak, using the exact peak position as the center of the region. As an additional analysis, one can also check whether individual particles are bound to their parent peak patch using a simple binding energy criterion such as:
\begin{equation}
\frac{1}{2}({\bf v}_p - {\bf v}_0)^2 + \phi({\bf x}_p - {\bf x}_0) < \phi({\bf x}_{\rm saddle} - {\bf x}_0)
\end{equation}
where $\phi({\bf x}_p - {\bf x}_0)$ is the particle potential energy with respect to the center defined by coordinate ${\bf x}_0$ and $\phi({\bf x}_{\rm saddle} - {\bf x}_0)$ is the potential energy at the most bound point on the saddle surface ${\bf x}_{\rm saddle}$.

\begin{figure}
\centering
\includegraphics[scale=0.17]{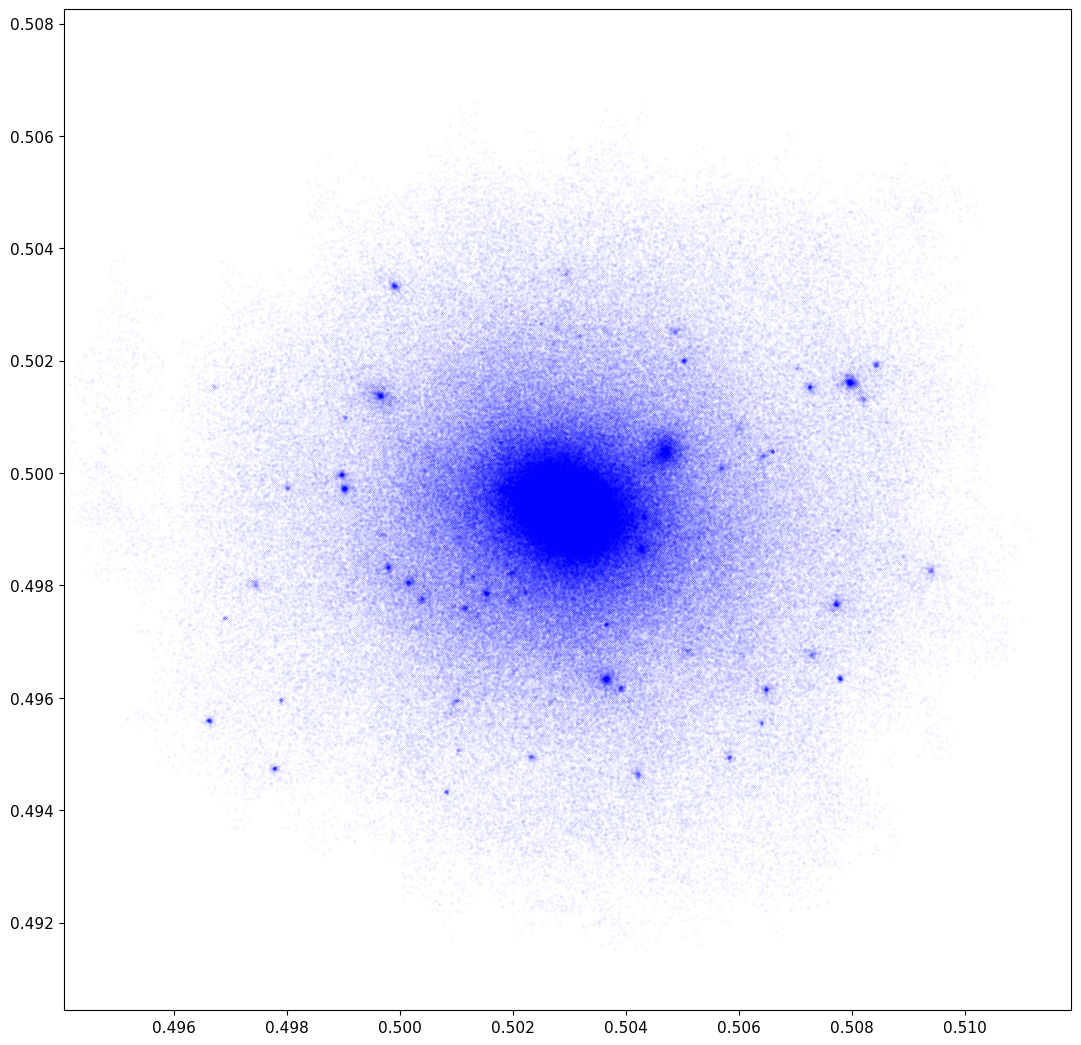}
\includegraphics[scale=0.17]{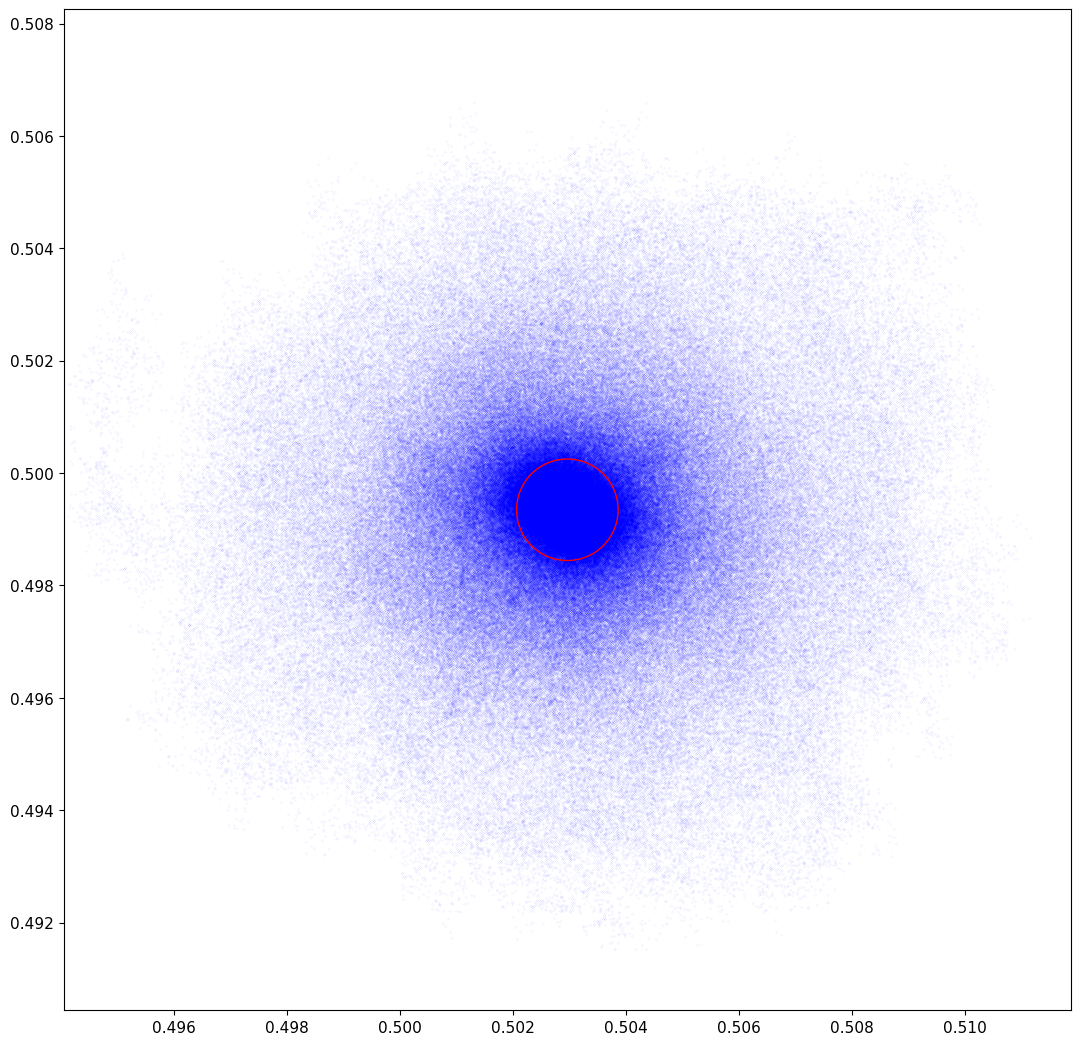}
\includegraphics[scale=0.17]{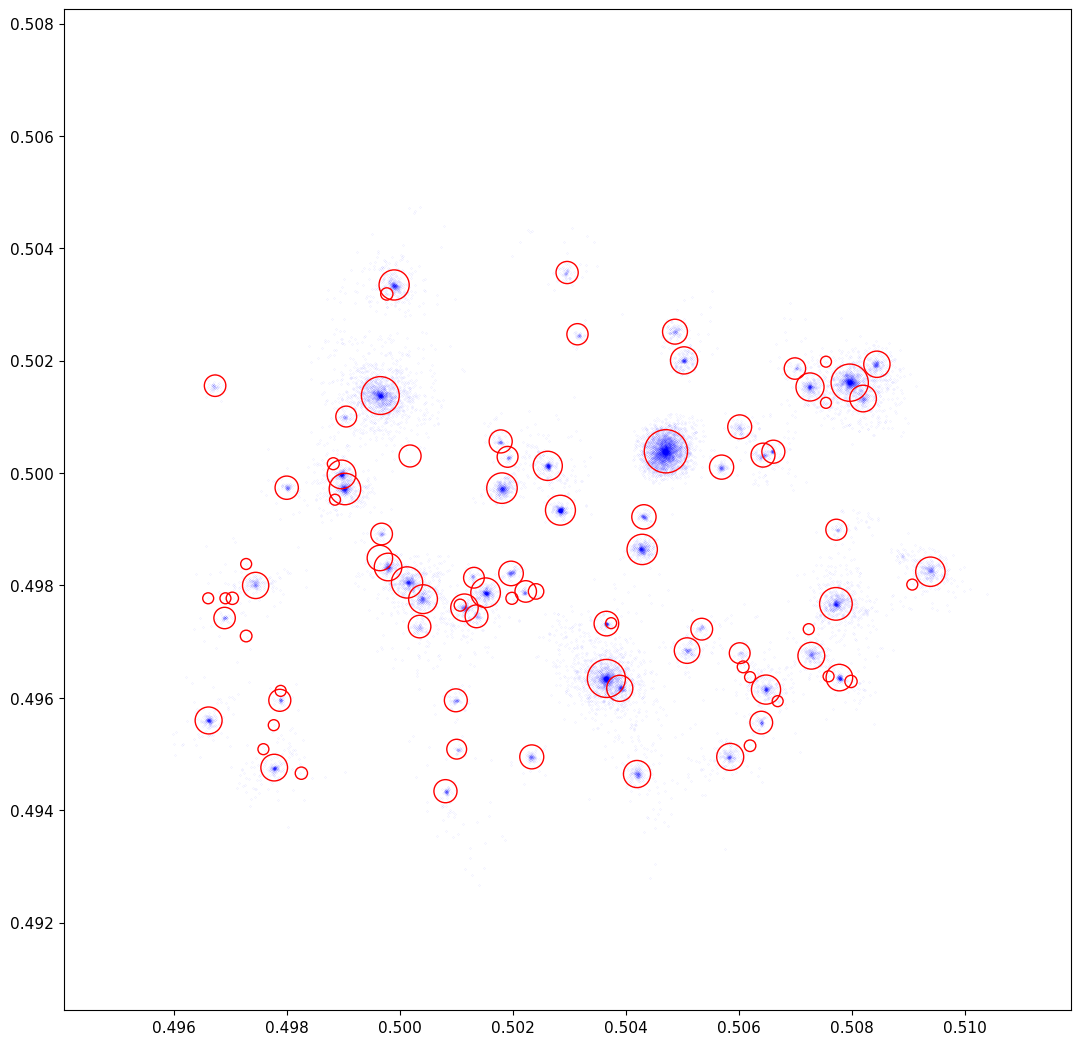}
\caption{Left panel: Particle positions of a zoom-in N-body simulation of a Milky Way size dark matter halo. Middle panel: Central peak of the dark matter halo (shown as a single large red circle) with particle (shown as blue dots) assigned to the central peak. Right panel: Satellite peaks of the dark matter halo (shown as the many small red circles) with particles (shown as blue dots) bound to each satellite. Credit figure and simulation: R. Teyssier.}
\label{fig:dmo_zoom}
\end{figure}

We see in Figure~\ref{fig:dmo_zoom} a typical example of such a central-satellite dark matter halo decomposition. The left panel shows all dark matter particles contained in the halo region, defined here in this particular example by $\rho_{\rm min}=80$. One can clearly see a large density peak centered on the image, with several smaller peaks scattered around the main peak. The right panel shows particles bound only to the smaller peaks, with the red circle dimensioned to match the mass of these satellite halos, also called sub-halos. The middle panel shows only the particles bound to the central peak, also called main halo. The red circle is also proportional to the mass of the main halo. This main halo versus sub-halos decomposition is crucial for the next step in our journey toward producing mock galaxy catalogs, as it helps us decompose the galaxy population into central galaxies and satellite galaxies, a crucial aspect of our understanding of the galaxy spatial distribution.

\subsection{Merger Trees and Galaxy Painting}

Once we know the statistics of dark matter halos, and thanks to accurate N-body simulations, the centers of main halos and sub-halos, we can try and populate this detailed halo population with galaxies. The main technique used in the current exploitation of large-scale galaxy surveys is the so-called {\it abundance matching} technique. The idea is surprisingly simple: if we know on one hand to exquisite precision the main halo mass function, as predicted by the theory and supported by N-body simulations, and on the other hand the stellar mass function of central galaxies, as observed and measured in various galaxy surveys, it is very tempting to match the halo mass of a halo population defined by its number density $n(M_{\rm vir})$ to the stellar mass of a galaxy population defined by its number density $n(M_*)$. Matching the 2 populations produces a fundamental quantity called the Stellar to Halo Mass Relation (SHMR) that we can use to populate central galaxies in our main halo catalogs produced by N-body simulations. A popular SHMR was proposed by \cite{Behroozi.2013} and takes the simple analytical form:
\begin{equation}
\frac{M_*}{M_h} = 2F \left[ \left( \frac{M_{\rm vir}}{M_1} \right)^{-\beta} + \left( \frac{M_{\rm vir}}{M_1} \right)^{\gamma} \right]^{-1}
\end{equation}
where the free parameters are the pivot mass $M_1 \simeq 10^{12}$~M$_\odot$, the stellar mass to halo mass fraction at the pivot mass $F \simeq$3.5\%, and the two slopes $\beta \simeq 1.4$ and $\gamma \simeq 0.6$. Trying to explain the origin of the SHMR is one of the main goal of galaxy formation theory. But we adopt here a pragmatic and empirical approach and only need this form as a way to assign the stellar mass of the galaxy at the center of its parent halo. 

This method has been shown to work beautifully when matching not only the stellar mass, but also any luminosity defined in any observational band associated to a given telescope. Once the number density of the halo population and the galaxy population are matched, the resulting spatial distribution also agree very well. At scales smaller than 1~Mpc, however, greater care must be taken on how to handle satellite galaxies. Indeed, around massive central galaxies bright satellite galaxies can be detected by current surveys and contribute to the measured clustering statistics. In this case, several simple recipes have been proposed to handle satellites in a similar way (Sub-Halo Abundance Matching or SHAM) \cite{Kravtsov.2004}. This requires unfortunately to develop a new technique called {\it merger trees} \cite{Tweed.2009,Behroozi.2013vab} to be able to follow main halos and sub-halos as a function of time, up to the point when and where a main halo becomes the sub-halo of a bigger main halo. 

Merger trees are very complex algorithms suffering from a lot of poorly understood limitations but these techniques have improved significantly in the past decade. We can now track accurately small main halos accreting into a larger main halo, monitoring how their mass is increasing steadily until it reaches a maximum usually just before or after merging into the main halo, and then steadily decreasing as the now satellite halo gets stripped by various physical or numerical effects. SHAM techniques traditionally compute the stellar mass or the luminosity of the satellite galaxies as a function of their maximum (main then sub-) halo mass, using the same SHMR than for central galaxies. This simple technique turns out to be also very accurate when trying to fit the correlation functions of galaxies even at small scales and down to small galaxies. 

More sophisticated techniques can take advantage of these merging tress. Regulator models try and predict the star formation rate using a simple stationary close-box model for galaxies where the galaxy gas mass obeys a simple ``bath tube model'':
\begin{equation}
\dot{M}_{\rm gas} = f_b \dot{M}_{\rm acc} - \dot{M}_*-\eta_{\rm wind}\dot{M}_* \simeq 0~~~{\rm so~that}~~~\dot{M}_* = \frac{f_b \dot{M}_{\rm acc}}{1+\eta_{\rm wind}}
\end{equation}
where $\eta_{\rm wind}$ is called the {\it mass loading factor} and models the gas mass entrained out of the galaxy by supernovae-driven winds, $f_b=\Omega_b/\Omega_m$ is the baryon fraction and $\dot{M}_{\rm acc}$ is the total matter accretion rate onto the halo, as produced by the merger tree algorithm. In the spirit of the abundance matching techniques, regulator models use a simple fit for the wind mass loading factor, such as the one proposed by \cite{Kravtsov.2022} to reproduce many properties of the low mass galaxy population:
\begin{equation}
\eta_{\rm wind}(M_{\rm vir})\simeq \left( \frac{M_{\rm vir}}{10^{12}M_\odot}\right)^{-1.5} 
\end{equation}
Finally, the most sophisticated approach based on merger trees are called {\it semi-analytical models} or SAM, and solve a complex set of ODE to describe the evolution of the full galaxy ecosystem, including different gas components (hot, warm, cold), stellar components (thin disk, thick disk, bulge) and a central supermassive black hole (SMBH). These models are much more physically motivated and are usually considered as a first and significant step towards galaxy formation physics. 

\subsection{Baryonification}

Before we embark in galaxy formation physics in the next Chapter, let's discuss an interesting numerical method that aims at modeling the effect of baryons on the matter distribution in the Universe, as observed for example in convergence maps obtained via weak gravitational lensing in large-scale galaxy surveys. This method is the culmination of decades of effort trying to estimate the back reaction of baryons to the dark matter distribution at small scales \cite{Rudd.2008,Guillet.2010gs}, effectively re-distributing matter from small scales to larger scales via feedback effects (mostly SMBH driven outflows) as well as from large scales to smaller scales via cooling and dissipation. The next effect was quite unclear at first, but advances in galaxy formation theory combined with detailed observations of the hot gas in galaxy groups and galaxy clusters indicate that matter on scale around 1 to 10 Mpc is significantly smoother than pure dark matter N-body simulations predict. Until galaxy formation theory is completely understood, we need to come up with intermediate, hopefully cheap solutions to model these uncertainties in the mass distribution at small scales. 

These techniques have received recently considerable attention because recent weak lensing results seen to favor strong feedback effects. These methods are called {\it baryonification methods} \cite{Schneider.2015,Schneider.2020rb8,Arico.2021}, as they allow one to modify the dark matter particle positions from N-body simulations to mimic the effect of baryonic physics. The technique is quite simple. It relies first on a proper main halo catalog where each halo is described by its best-fit NFW profile. In a second step, a simple halo-based spherical model for the gas distribution is added via an analytical profile. This gas profile is parametrized by a few free parameters. In a third step, a central galaxy with stellar mass set by abundance matching is placed at the center. In a fourth step, the response of dark matter to this modification of the halo mass distribution is computed via standard adiabatic expansion/contraction of dark matter particle orbits. Finally, all these different components (central galaxy, extended gas distribution, adiabatically contacted/expanded dark matter distribution) are combined to create a new {\it baryonified} mass distribution. The difference between the initial NFW mass profile and the final baryonified mass profile is used to construct a {\it displacement field} that is applied to the N-body particles, which concludes the baryonification of the original N-body simulations.

\subsection{Light Cones}

\begin{figure}
\centering
\includegraphics[scale=0.4]{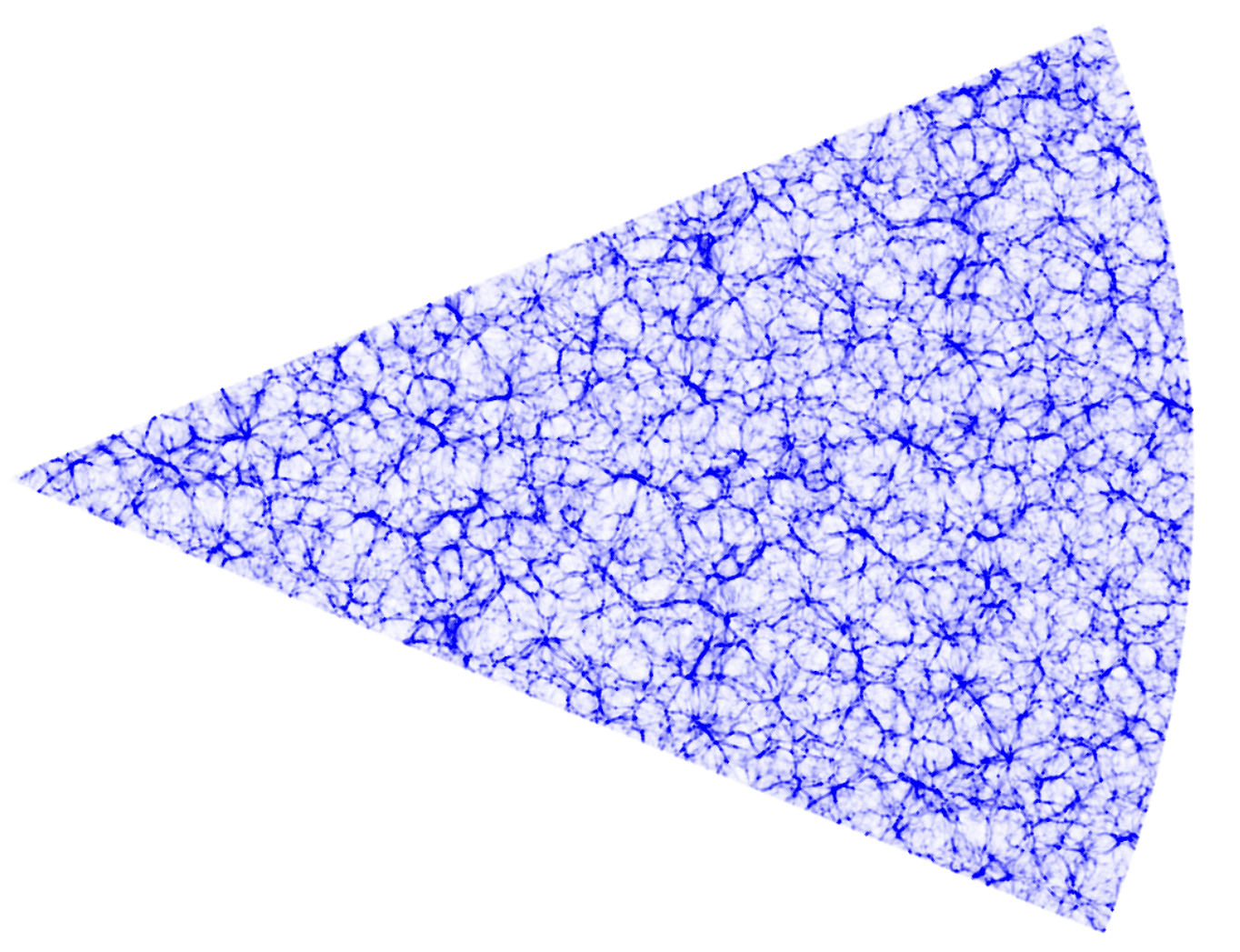}
\includegraphics[scale=0.2]{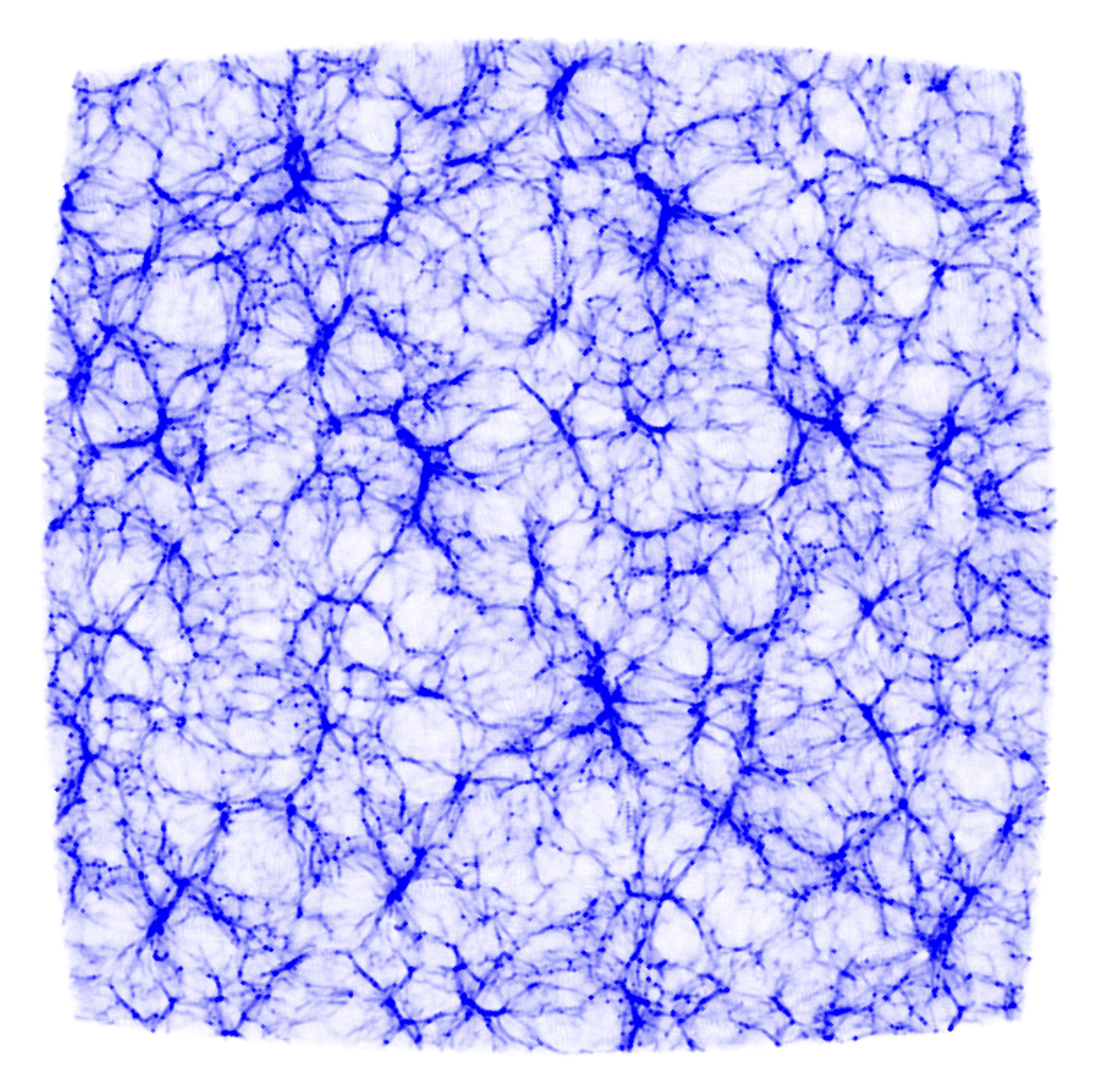}
\caption{Left panel: Light cone as seen from an observer sitting at the tip of the triangle within a thin slice on the sky and up to redshift $z=0.5$. Right panel: Light cone as seen from the same observer but face-on for a thin slice in redshift around redshift $z=0.25$. Credit figure and simulation: T. Akieda-Codron.}
\label{fig:lightcone}
\end{figure}

The final products that can be easily produced via N-body simulations are light cones \cite{Fosalba.2014}. An observer is chosen at any preferred location in the computational box and a photon spherical shell is shot at large distance corresponding to the starting redshift of the light cone. This photon shell moves and converges at the speed of light towards the observer and collect dark matter particles, main halos and sub-halos as they are swept up by the photon sphere. At the end of the simulation, a large data set containing all the necessary information is stored using a simple comoving coordinate system where the observer is at the origin and particles at larger distances carry information at larger redshift. Figure~\ref{fig:lightcone} shows such a light cone generated by small N-body simulation. The side-on view reveals the typical filamentary structure of the Universe. The face-on view shows a typical field of view a few degrees aside realistically mocking present and upcoming galaxy surveys.

%% file: chapter3.tex
\section{Hydrodynamics and Galaxy Formation}

\subsection{Introduction}

Modeling the gas component requires very different numerical tools. This comes from a fundamental different properties of baryons: they are highly collisional. In plasma physics, collisional fluids are described using the Boltzmann equation that writes \cite{Mihalas.1984}:
\begin{equation}
\frac{\partial f}{\partial t} + {\bm v} \cdot \frac{\partial f}{\partial {\bm x}} + {\bm g} \cdot \frac{\partial f}{\partial {\bm v}}
= \int_{4\pi} \int_{\mathbb{R}^3} \left(f_1' f_2' - f_1 f_2\right) \sigma \left|{\bm v}_1-{\bm v}_2  \right| \, {\rm d}\Omega \, {\rm d}^3 v_2
\label{eq:boltzman}
\end{equation}
The left-hand side is identical to the Vlasov equation and expresses particle number conservation in phase-space. The right-hand side, however, was zero for dark matter. For baryons, it is not and represents the effect of binary collision. It is called the {\it collision integral}. Because an elastic binary collision conserves the mass, momentum and energy of the two collision partners, one can multiply the Boltzmann equation by $m$, $m{\bm v}$ and $\frac{1}{2}m v^2$ and integrate over velocity space, the contribution of the collision integral will vanish. This leads to exact moment's equations called the conservation of mass:
\begin{equation}
\frac{\partial}{\partial t} \rho + {\bm \nabla} \cdot (\rho {\bm u}) = 0
\end{equation}
the conservation of momentum:
\begin{equation}
\frac{\partial}{\partial t} (\rho \, {\bm u}) + {\bm \nabla} \cdot (\rho \, {\bm u} \otimes {\bm u} + \mathbb{P}) = \rho \, {\bm g}
\end{equation}
and finally the conservation of energy:
\begin{equation}
\frac{\partial E}{\partial t} + {\bm \nabla} \cdot (E \, {\bm u} + \mathbb{P} {\bm u} + {\bm Q}) = \rho \, {\bm g} \cdot {\bm u}
\end{equation}
We have introduced multiple key fluid quantities that we now define properly.
The fluid mass density $\rho({\bm x},t)$ has been already defined for dark matter in the right-hand side of Poisson equation. This is now the gas density defined as the zero-th order moment of the distribution function $f({\bm x}, {\bm v}, t)$. We also introduce the fluid velocity ${\bm u}({\bm x},t)$ as:
\begin{equation}
\rho({\bm x},t) = \int_{\mathbb{R}^3} m f{\rm d}v^3~~~{\rm and}~~~ \rho({\bm x},t) {\bm u}({\bm x},t) =  \int_{\mathbb{R}^3} m {\bm v}f{\rm d}v^3
\end{equation}
The fluid velocity here is fundamentally different from the particle velocity. The fluid velocity ${\bm u}$, also called the bulk velocity, is the average velocity of particles within the same volume element ${\rm d}^3 x$, while the particle velocity ${\bm v}$ stands for individual particle's velocity. This allows us to define two different regimes, the macroscopic regime for the fluid and the microscopic regime for individual particles. This also corresponds to a scale separation between microscopic scales, below the {\it particle mean free path} and macroscopic scales, above the particle mean free path defined later. We can also define the relative velocity ${\bm w} = {\bm v} - {\bm u}$, also known as the thermal velocity. 
This allows us to define the {\it pressure tensor} as the 3x3 matrix:
\begin{equation}
P_{ij} = \int_{\mathbb{R}^3} m \, w_i w_j f \, \dd^3 v
\end{equation}
and the {\it heat flux} as the vector:
\begin{equation}
\mathbf{Q} = \int_{\mathbb{R}^3} m \frac{1}{2} w^2 {\bm w} f \, \dd^3 v
\end{equation}
Finally, we define the macroscopic or fluid total energy $E$ as:
\begin{equation}
E({\bm x},t) = \int_{\mathbb{R}^3} \frac{1}{2}m v^2 f{\rm d}v^3 = \frac{1}{2}\rho {\bm u}^2 + e({\bm x},t)
\end{equation}
where the internal, microscopic or thermal energy $e$ is defined as:
\begin{equation}
e = \int_{\mathbb{R}^3} \frac{1}{2} m w^2 f \, \dd^3 v = \frac{1}{2} \mathrm{Tr}(\mathbb{P})
\end{equation}

Let's go back and scrutinize the collision integral that stands as the right-hand side of Boltzmann's equation. The important quantity in there is the interaction cross section $\sigma({\bm \Omega}, \left| {\bm v}_1-{\bm v}_2\right|)$. A simple model for collision is the so-called {\it hard sphere model}, for which $\sigma = \sigma_0$ where $\sigma_0 \simeq 10^{-15}~{\rm cm}^2$ is related to the typical radius of atoms and molecules in the fluid via simple geometry $\sigma_0 = \pi r_0^2$. One can define the collision rate using the moment of the collision integral (using only outgoing collisions) as:
\begin{equation}
C_{\rm coll} = \int_{\mathbb{R}^3}\int_{\mathbb{R}^3}\int_{4\pi} f_1 f_2 \, \sigma  \left|{\bm v}_1-{\bm v}_2  \right| \, {\rm d}^3 v_1 \, {\rm d}^3 v_2 \, {\rm d}\Omega
\end{equation}
If the collision rate is high enough, and we will compute it in a moment, the distribution function $f$ is driven towards the equilibrium distribution function $f_0$ for which the collision integral vanishes. This equilibrium distribution function is the Maxwellian (Gaussian) distribution:
\begin{equation}
f_0({\bm v}) = \frac{\rho}{m}\frac{1}{\big(2\pi a^2 \big)^{3/2}} \exp \left( - \frac{1}{2} \frac{\big({\bm v}-{\bm u}\big)^2}{a^2} \right)
\end{equation}
where $a$ is the one-dimensional microscopic velocity dispersion, also called the isothermal sound speed. It is used to define the gas temperature $T$ as:
\begin{equation}
a^2 = \frac{k_B T}{m}
\end{equation}
Because the Maxwellian is isotropic and symmetric, the fluid equations we derived earlier simplifies greatly as the pressure tensor $\mathbb{P} = P \mathbb{I}$ becomes diagonal and the heat flux vanishes ${\bm Q}=0$. In this simplified form, the fluid equations become the Euler equations. 
Note that for a Maxwellian gas, we have the relation $P = \frac{2}{3}e$, which corresponds to what the traditional definition of an ideal gas with adiabatic exponent $\gamma=\frac{5}{3}$. We also obtain the classical ideal gas Equation-Of-State (EOS):
\begin{equation}
P = \frac{\rho k_B T}{m}~~~{\rm and}~~~e = \frac{3}{2} \frac{\rho k_B T}{m}
\end{equation}
If one uses the previous definition for the collision rate, injection $f_0$ for the distribution function $f_1$ and $f_2$ and using the binary collision invariants (mass, momentum and energy), one gets for hard spheres:
\begin{equation}
C_{\rm coll} \simeq n^2 \sigma_0 \sqrt{\frac{k_B T }{m}}
\end{equation}
The particle number density $n=\rho/m$ allows us to compute the typical time and distance between collisions as:
\begin{equation}
\tau_{\rm coll} = \frac{n}{C_{\rm coll}} \simeq \frac{1}{n \sigma_0 \sqrt{\frac{k_B T}{m}}}~~~{\rm and}~~~
\lambda_{\rm coll} =  \sqrt{\frac{k_B T}{m}} \tau_{\rm coll} \simeq \frac{1}{n \sigma_0}
\end{equation}
The regime for which $f \simeq f_0$ is a valid assumption is called Local Thermodynamical Equilibrium (LTE). If one defines traditionally the typical length scale of the macroscopic system as the pressure scale height:
\begin{equation}  
\frac{P}{H_P} = \frac{\partial P}{\partial x},
\end{equation}
the validity of the LTE approximation can be written as $H_P \gg \lambda_{\rm coll}$. In this limit, the fluid is strongly collisional and follows the Euler-Poisson equation, summarized here as:
\begin{align}
&\frac{\partial \rho}{\partial t} + {\bm \nabla} \cdot (\rho {\bm u}) = 0 \\
&\frac{\partial}{\partial t} (\rho {\bm u}) + {\nabla} \cdot (\rho {\bm u} \otimes {\bm u} + P\mathbb{I} ) = \rho {\bm g} \\
&\frac{\partial E}{\partial t} + {\nabla} \cdot (E+P) {\bm u} = \rho {\bm g} \cdot {\bm u} \\
& \Delta \phi = 4\pi G \rho~~~{\rm and}~~~{\bm g} = - {\bm \nabla}\phi
\end{align}
If the system approches the weakly collisional limit $H_P \simeq \lambda_{\rm coll}$, one can derive, using the so-called Chapman-Enskog expansion \cite{Mihalas.1984}, an accurate form for the pressure tensor:
\begin{equation}
{\mathbb P} = P {\mathbb I}
-
\mu \left(
{\mathbb G} + {\mathbb G}^{\rm T}
- \frac{2}{3} (\nabla \cdot {\bm u}) {\mathbb I}
\right)~~~{\rm with}~~~ G_{ij} = \frac{\partial u_i}{\partial x_j}
\end{equation}
where $\mu$ is the viscosity coefficient, that the Chapman-Enskog theory predicts to be
\begin{equation}
\mu = \rho \lambda_{\rm coll} \sqrt{\frac{k_B T}{m}} = \rho \nu_{\rm coll}
\end{equation}
and also for the heat flux:
\begin{equation}
\mathbf{Q} = - \kappa \nabla T
\end{equation}
where we introduce the heat conduction coefficient $\kappa$, which, according to the Chapman-Enskog theory,
takes the value
\begin{equation}
\kappa = \rho \lambda_{\rm coll} \sqrt{\frac{k_B T}{m}} \frac{k_B}{m} = \rho \frac{k_B}{m} \nu_{\rm coll}
\end{equation}
where we defined the microscopic diffusion coefficient:
\begin{equation}
\nu_{\rm coll} = \lambda_{\rm coll} \sqrt{\frac{k_B T}{m}}
\end{equation}
In the limit $H_P \ll \lambda_{\rm coll}$, we are in the collisionless limit, where the collision integral can be ignored and we are back to Vlasov-Poisson equations. The previous derivations are based on the simplified hard sphere model. In case of a plasma with charged particles, one has to use the Coulomb interaction and follow Spitzer's theory to derive all corresponding terms. Our main conclusions still apply.

\subsection{Euler-Poisson using Grids}

We now discuss numerical methods to solve the Euler-Poisson equations using grids. We adopt the so-called Finite Volume (FV) approach, and for sake of simplicity our discussion is restricted to the one-dimensional case. We discretize space into volume elements defined as:
\begin{equation}
V_i = \left[ x_{i-1/2}, x_{i+1/2}\right]
\end{equation}
with a constant mesh size $\Delta x = x_{i+1/2}-x_{i-1/2}$. We adopt the FV discretization of the fluid at some time step $t^n$, define the cell-averaged conservative variables as:
\begin{equation}
{\bm U}_i^n = \frac{1}{\Delta x} \int_{x_{i-1/2}}^{x_{i+1/2}} {\bm U}(x,t^n) {\rm d}x
\end{equation}
where the vector of conservative variable is defined as ${\bf U}=\left( \rho, \rho u, E \right)$.
Using the divergence theorem, we can write an exact conservative update of these conservative variables as:
\begin{equation}
{\bm U}_i^{n+1} = {\bm U}_i^n - \frac{\Delta t}{\Delta x} \left( {\bm F}_{i+1/2}^{n+1/2} - {\bm F}_{i-1/2}^{n+1/2}\right)
\end{equation}
where the vector of fluxes is defined as ${\bm F} = \left( \rho u, \rho u^2 + P, (E+P)u \right)$ and the time averaged flux at interface $x_{i+1/2}$ is defined as:
\begin{equation}
{\bm F}_{i+1/2}^{n+1/2} = \frac{1}{\Delta t} \int_{t^n}^{t^{n+1}} {\bm F}(x_{i+1/2},t) {\rm d}t
\end{equation}
The difficulty here, like for the phase-space Vlasov-Poisson solver, is to compute the time averaged interface fluxes. Godunov's method recommends the use of properly upwinded fluxes, depending on the sign of the speed of each waves. If we consider the simple case of the advection equation, for which the velocity $u$ is a prescribed constant $a$, we have:
\begin{equation}
\rho_i^{n+1} = \rho_i^n - \frac{\Delta t}{\Delta x} \left( f_{i+1/2}^{n+1/2} - f_{i-1/2}^{n+1/2} \right)
\end{equation}
with interface fluxes defined using the upwind solution:
\begin{align}
& {\rm if}~~a>0,~~~ f_{i+1/2}^{n+1/2}=a \rho_i^n \\
& {\rm if}~~a<0,~~~ f_{i+1/2}^{n+1/2}=a \rho_{i+1}^n
\end{align}
Upwinding guarantees the positivity and the convergence of the solution. Indeed, in the case $a>0$, we obtain the following fully discrete scheme:
\begin{equation}
\rho_i^{n+1} = \rho_i^n \left( 1 -C \right) + \rho_{i-1}^n C
\end{equation}
if the Courant-Friedrich-Levy (CFL) number $C$ satisfies:
\begin{equation}
C= a \frac{\Delta t}{\Delta x} < 1
\end{equation}
We can Taylor expand the solution both in space and time as:
\begin{align}
&\rho_{i-1}^n = \rho_i^n - \Delta x\frac{\partial \rho}{\partial x} + \frac{\Delta x^2}{2} \frac{\partial^2 \rho}{\partial x^2}+{\cal O}(\Delta x^3) \\
&\rho_{i}^{n+1} = \rho_i^n + \Delta t\frac{\partial \rho}{\partial t} + \frac{\Delta t^2}{2} \frac{\partial^2 \rho}{\partial t^2}+{\cal O}(\Delta t^3)
\end{align}
and after some manipulations obtain the modified equation of the numerical scheme as:
\begin{equation}
\frac{\partial \rho}{\partial t} + a \frac{\partial \rho}{\partial x} = \frac{a \Delta x}{2}(1-C) \frac{\partial^2 \rho}{\partial x^2} + {\cal O}(\Delta x^2)
\end{equation}
Interestingly, we see that the numerical scheme does not solve for the original advection scheme, for which the right-hand side of the previous equation should be zero, but a modified equation featuring a diffusion term, for which the numerical diffusion coefficient reads:
\begin{equation}
\nu_{\rm num} = \frac{a \Delta x}{2} (1-C)
\end{equation}
Here again, the CFL condition $C<1$ guarantees that the diffusion coefficient is positive, a testament to the stability of the numerical scheme.
We also see that the scheme is second-order accurate for a single time step and only first-order accurate overall after integration up to some final time $T$.

For the Euler equations, things get more complicated, as we don't have a single advection velocity, but we have now 3 distinct waves speed $u-c_s$, $u$ and $u+c_s$, where we define the adiabatic sound speed as:
\begin{equation}
c_s = \sqrt{\frac{\gamma P}{\rho}}
\end{equation}
The strategy is here to use a Riemann solver in order to perform the proper wave-by-wave upwinding \cite{Toro.1999}. We can write formally:
\begin{equation}
{\bm F}_{i+1/2}^{n+1/2} = {\rm RP}({\bm U}_{i}^n, {\bm U}_{i+1}^n)
\end{equation}
where ${\rm RP}({\bm U}_L,{\bm U}_R)$ stands for the solution to the Riemann problem defined by the left and right piecewise constant initial states as arguments. One can see this Riemann solver as a black box performing the proper nonlinear wave decomposition of the Riemann solution and delivering properly upwinded fluxes. In this case, the CFL condition and the numerical diffusion coefficient write:
 \begin{equation}
C= (\left| u \right| + c_s) \frac{\Delta t}{\Delta x} < 1~~~{\rm and}~~~\nu_{\rm num} = \frac{(\left| u \right| + c_s) \Delta x}{2} (1-C)
\end{equation}
It is possible to implement higher-order versions of the Godunov scheme, for which the numerical diffusion is considerably smaller. The MUSCL-Hancock scheme is a particularly efficient second-order scheme. Developing very high order scheme is a topic of intense research these days.

We also need to compute the gravitational acceleration, using any of the field solvers discussed in the first Chapter. The gravity is traditionally added as a source term using the so-called operator split approach, where one solves first for the Euler equations without gravity, and then one modifies the new momentum and total energy using the gravitational acceleration computed at time $t^n$.

A key feature of Godunov schemes, or any modern grid-based schemes, is that they are strictly conservative for the fluid mass, momentum and total energy. This is particularly important to handle strong shock waves, a key physical process in galaxy formation that we will describe in the next sections.

\subsection{Hydrostatic Equilibrium}

We now describe the physics of baryons in the context of hierarchical structure formation. We adopt here also a halo-centric view, in which we assume that an overdensity $\delta_{\rm ini}$ of initial comoving radius $R$ containing a mass $M=\frac{4\pi}{3}\rho_m R^3$ will collapse before today if its linearly extrapolated overdensity satisfies $\delta > \delta_c$. We can use the spherical collapse model to model the dynamics of each spherical shell, following first the expansion, reaching its turn-around radius and finally collapsing to its final virialized state \cite{Gunn.1972}. 

The main difference with dark matter is that baryons are collisional, so there will be no shell crossing and violent relaxation. Instead, just before collapse, a shock wave propagating from the inside out will halt the collapse and transfer the kinetic energy of the free-falling gas shell into thermal energy. In the frame of the shock, called in this context the accretion shock, we can write the conservation of mass, momentum and energy across the shock discontinuity as:
\begin{align}
\rho_1 v_1 & =  \rho_2 v_2 \\
\rho_1 v_1^2 + P_1 & = \rho_2 v_2^2 + P_2\\
\left( \frac{1}{2}\rho_1 v_1^2 + \frac{\gamma}{\gamma-1}P_1 \right) v_1 & = \left( \frac{1}{2}\rho_2 v_2^2 + \frac{\gamma}{\gamma-1}P_2 \right) v_2 
\end{align}
These equations, called the Rankine-Hugoniot relations, can be solved easily in the case of a strong accretion shock \cite{Toro.1999}. We assume first that the velocity in the virialized region is zero. The halo is in hydrostatic equilibrium. We also assume that the gas pressure is negligible outside the halo, when baryons are free-falling and collapsing towards the halo central region. Injecting the first equation into the third one, we obtain immediately the post-shock (or virial) temperature as a function of the pre-shock velocity as:
\begin{equation}
\frac{k_B T_{\rm vir}}{m_H} = \frac{\gamma-1}{2\gamma}v_{\rm collapse}^2~~~{\rm with}~~~v_{\rm collapse}^2 \simeq \frac{G M}{R_{\rm vir}} \simeq V_{\rm circ}^2
\end{equation}
where the collapse velocity can be computed using the spherical collapse model at virialization. Baryons do not reach virial equilibrium via relaxation of their orbits like dark matter, but via shocks. The collapse velocity depends on the exact collapse time of each spherical shell, which usually results in a complex temperature profile. 

\begin{figure}
\centering
\includegraphics[scale=0.5]{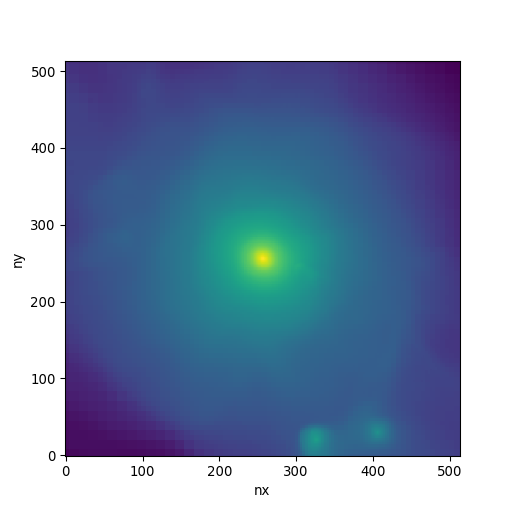}
\includegraphics[scale=0.5]{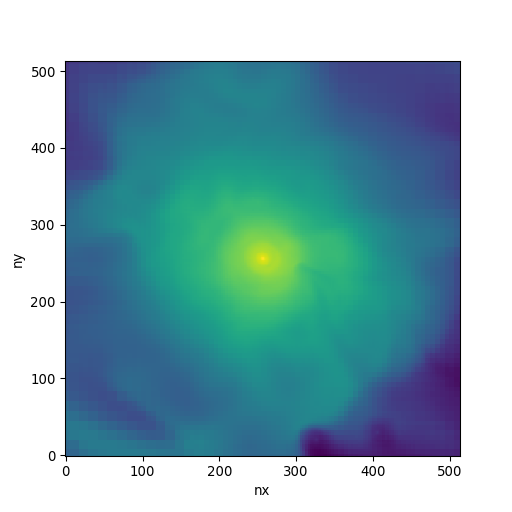}
\caption{Left panel: gas density of the halo simulated using only non-radiative physics in log-scale. Right panel: corresponding gas temperature also in log scale. Credit figure and simulation: R. Teyssier.}
\label{fig:nonrad}
\end{figure}

Numerical simulations can be used to estimate the exact structure of the gas inside virialized halos. These simulations are called {\it non-radiative} because they only include dark matter dynamics and the Euler-Poisson equations, so that mass, momentum and energy are conserved. We show in Figure~\ref{fig:nonrad} the density and temperature maps of such a non-radiative simulation in a cosmological zoom-in of a Milky-Way-size halo. The radius of the halo is roughly 250~kpc and images are 2 virial radii across. We see that the gas density is centrally concentrated and the temperature is not a constant, but follows also a centrally concentrated distribution. 

\begin{figure}
\centering
\includegraphics[scale=0.25]{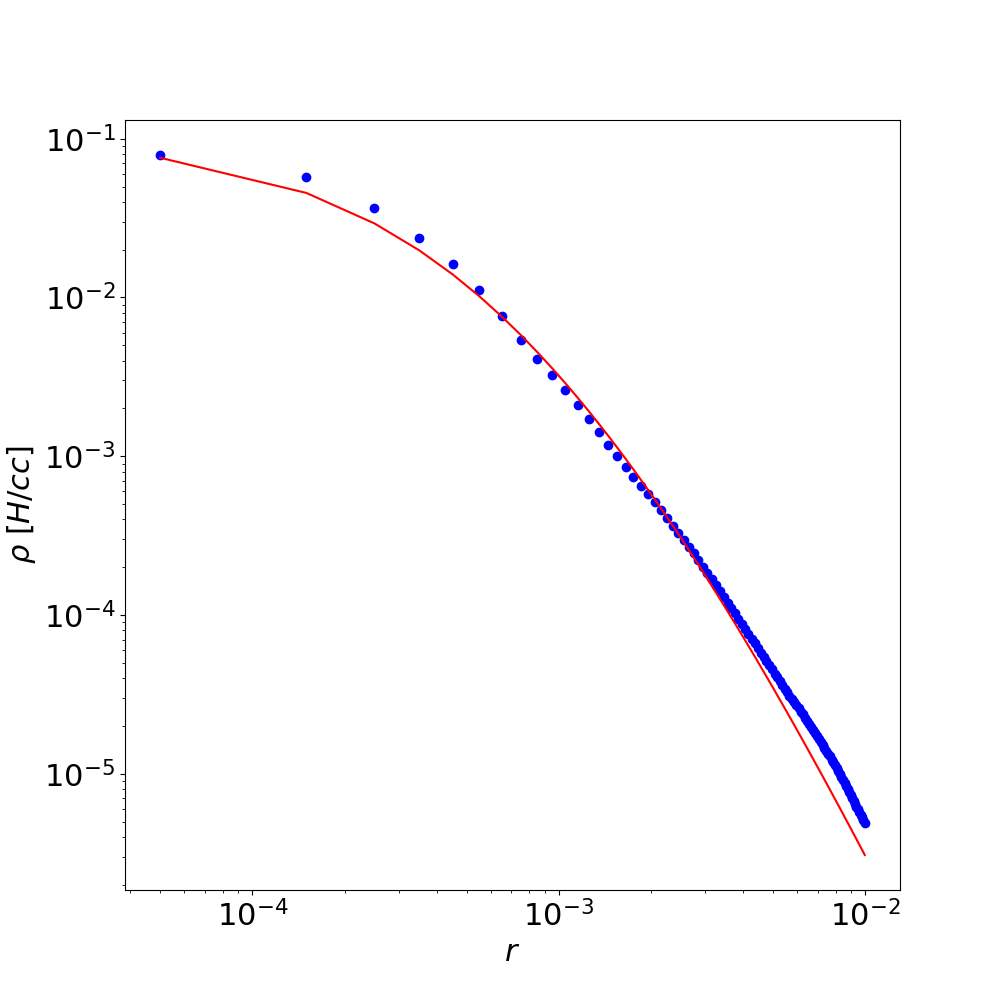}
\includegraphics[scale=0.25]{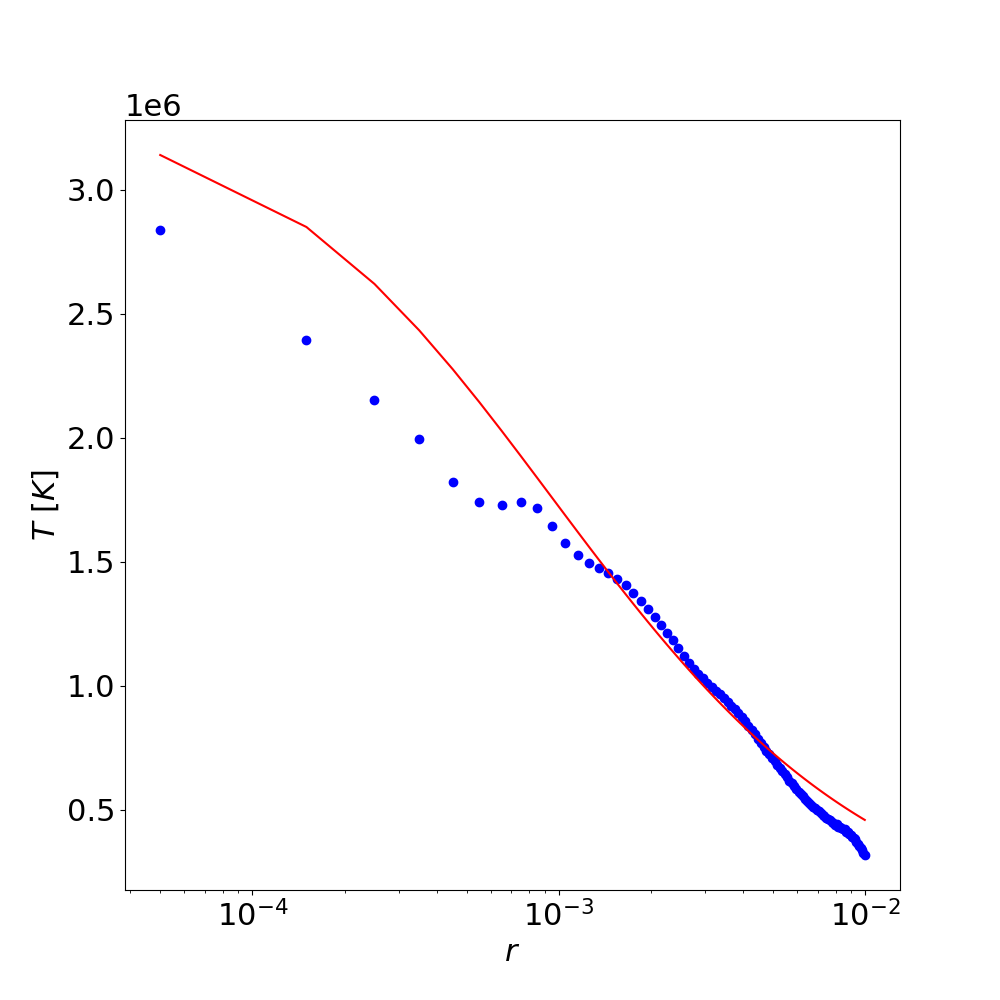}
\caption{Left panel: gas density profile of the simulated halo in the non-radiative run. Blue dots represents the simulation data, while the red solid line show the analytical model. Right panel: Same but for the gas temperature. Credit figure and simulation: R. Teyssier.}
\label{fig:hydrostaticfit}
\end{figure}

Using the equation of hydrostatic equilibrium, we can try and compute analytically these distributions \cite{Komatsu.2001}. We use the momentum conservation equation, assuming that the gas is static ${\bm v}$=0:
\begin{equation}
\frac{1}{\rho} \frac{\partial P}{\partial r} = - \frac{G M}{r^2}
\end{equation}
where we assume that the halo follows the NFW spherical profile. We also assume that the gas in the halo follows a polytropic relation with:
\begin{equation}
P = P_0 \left( \frac{\rho}{\rho_0} \right)^{\Gamma}
\end{equation}
where $\Gamma$ is the polytropic index and $\rho_0$ and $P_0$ are the central density and pressure. We can solve the previous ordinary differential equation using the boundary condition $P \rightarrow 0$ as $r \rightarrow +\infty$ and obtain:
\begin{equation}
T(r) = T_0 \frac{\ln{1+x}}{x}~~~{\rm and}~~~\rho(r) = \rho_0 \left( \frac{\ln{1+x}}{x} \right)^{\frac{1}{\Gamma-1}}~~~{\rm with}~~~x = \frac{r}{r_s}
\end{equation}
For typical cosmological and halo parameters, the constants $T_0$ and $\rho_0$ are given by:
\begin{equation}
\frac{k_B T_0}{m_H} = 4\pi G \rho_s r_s^2 \frac{\Gamma-1}{\Gamma}~~~{\rm and}~~~\rho_0 \simeq 0.2 \rho_s
\end{equation}
The only adjustable parameter of the model, once we have determined the NFW parameters $\rho_s$ and $r_s$ is the polytropic index $\Gamma$. We show in Figure~\ref{fig:hydrostaticfit} the comparison between the simulation and the analytical model with $\Gamma=1.19$ \cite{Ascasibar.2003}. The agreement is quite good, although far from perfect. This demonstrates that the non-radiative evolution of cosmological halos is relatively well understood, with simple physics featuring gravity, shock heating and the assembly of stable hydrostatic gas and dark matter halos.

One reason the agreement between the simulation and our simple analytical model is the fact that the gas is not in strict static thermal (or pressure) equilibrium. Some residual gas motions persist after the accretion shock, mostly in the outskirts of the halo \cite{Rabold.2017}. These residual motions are called turbulent motions. They can be modeled statistically like a random process providing turbulent pressure support. This extra term, sometime referred to as ``non-thermal pressure'' in cosmology, can contribute up to 10\% of the total pressure in galaxy clusters. Turbulence is probably the main contributor, but other sources of pressure support exist like magnetic fields or cosmic rays.

\subsection{Radiative Cooling}

One fundamental process we have ignored so far is gas cooling \cite{Rees.1977}. We have considered so far elastic collisions as the main factor leading the baryonic distribution function towards its LTE Maxwellian shape. Elastic collisions are also responsible for shocks, converting collapse kinetic energy into equilibrium thermal pressure support. A small fraction of binary collisions are in fact inelastic, in the sense that they do not conserve mass and energy anymore, because they emit a photon \cite{Sutherland.1993}. A typical example is collisional recombination, for which a free electron recombines with a ionized Hydrogen atom to form a neutral atom with its bound electron, emitting in the process a UV photon with energy equal to the binding energy of the electron. This reaction writes:
\begin{equation}
{\rm e}^- + {\rm H}^+ \rightarrow {\rm H}^0 + \gamma
\end{equation}
where $\gamma$ represent a photon of energy $h\nu_0=$13.6~eV. Assuming that the probability of a recombining collision is around $P_{\rm rec} \simeq 1\%$ and that the collision cross-section is given by the hard sphere model, we can compute the corresponding cooling rate as:
\begin{equation}
Q_{\rm rec} \simeq n_{\rm e^-} n_{\rm H^+} \sigma_0 \sqrt{\frac{k_B T}{m_H}}P_{\rm rec}h\nu_0 = n_H^2 \Lambda(T)
\end{equation}
where we have introduced the {\it cooling function} in units of $[ {\rm erg}~{\rm s}^{-1}{\rm cm}^3]$ that writes for recombination cooling:
\begin{equation}
\Lambda(T) \simeq x_{\rm e}^2 \sigma_0 \sqrt{\frac{k_B T}{m_H}}P_{\rm rec}h\nu_0
\end{equation}
where we have introduced the ionisation fraction $x_e(T)=n_{\rm e^-}/n_H$ and we have assumed charge neutrality $n_{\rm e^-} \simeq n_{\rm H^+}$. In most situations, the gas is fully ionized $x_e \simeq 1$ when $T>10^4$~K and becomes neutral  $x_e \simeq 0$ when $T<10^4$~K. Assuming $\sigma_0 \simeq 10^{-15}~{\rm cm}^2$, we find the classical result:
\begin{equation}
\Lambda(T) \simeq 10^{-22} \sqrt{\frac{T}{10^4~K}}~{\rm erg~s^{-1}cm}^3
\end{equation}
This simplistic derivation is actually quite accurate to estimate the cooling rate in the post-shock gas in the halo. In order to estimate how important is cooling for the gas evolution in the halo, one traditionally compute the cooling time as the ratio of the gas internal energy to the cooling rate:
\begin{equation}
t_{\rm cool} = \frac{\frac{3}{2}n_H k_B T_{\rm vir}}{Q_{\rm rec}(T_{\rm vir})} \simeq 100 \left( \frac{T_{\rm vir}}{10^4~K} \right)^{1/2} \left( \frac{n_H}{10^{-5}~{\rm cm}^{-2}}\right)^{-1}~{\rm Myr}
\end{equation}
For the gas density, we use $n_H \simeq 10^{-5}$H/cc, which corresponds to the mean halo overdensity $200 \rho_b$ where $\rho_b = \Omega_b \rho_c$ is the critical density multiplied by the baryon fraction. 

\begin{figure}
\centering
\includegraphics[scale=0.34]{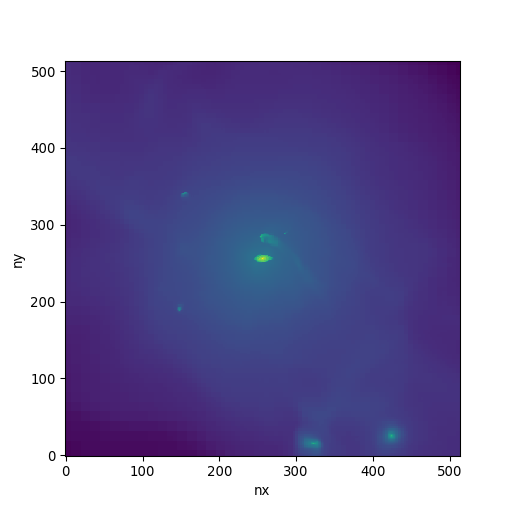}
\includegraphics[scale=0.34]{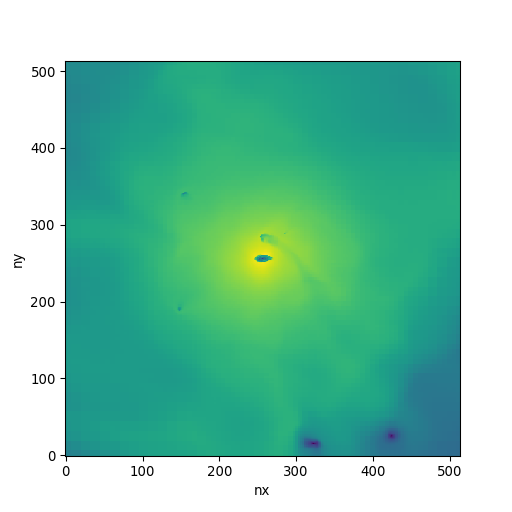}
\includegraphics[scale=0.34]{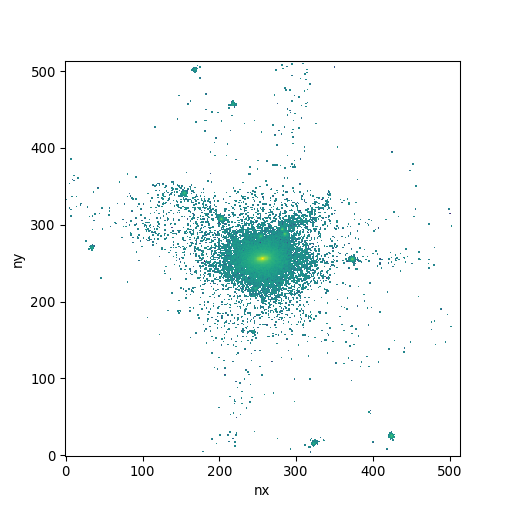}
\caption{Left panel: projected density map of a zoom-in cosmological simulation of a Milky-Way-size halo with radiative cooling and no feedback. Middle panel: projected temperature map. Right panel: stellar surface density map. The image is 500~kpc across. Credit figure and simulation: R. Teyssier.}
\label{fig:nofbkrun}
\end{figure}

For $T_{\rm vir}<10^4$~K, the gas is neutral and atomic cooling is absent. For small mass halos with $T_{\rm vir} \simeq 10^4$~K, the cooling time is very short, around 100~Myr, much smaller than the Hubble time of 13~Gyr, or the halo dynamical time $t_{\rm dyn} = R_{\rm vir}/V_{\rm circ} \simeq 1$~Gyr, interestingly also independent of the halo mass. For Milky-Way-size halos, we have $T_{\rm vir} \simeq 10^6$~K and the cooling time increases to around 1~Gyr, now comparable to the halo dynamical time. Finally, for large galaxy clusters with $T_{\rm vir} \simeq 10^8$~K, the cooling time is much longer than the halo dynamical time with $t_{\rm cool} \simeq 10$~Gyr, comparable to the Hubble time. 

\begin{figure}
\centering
\includegraphics[scale=0.34]{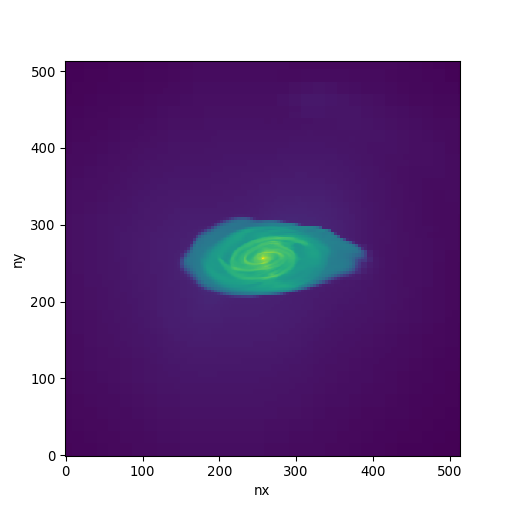}
\includegraphics[scale=0.34]{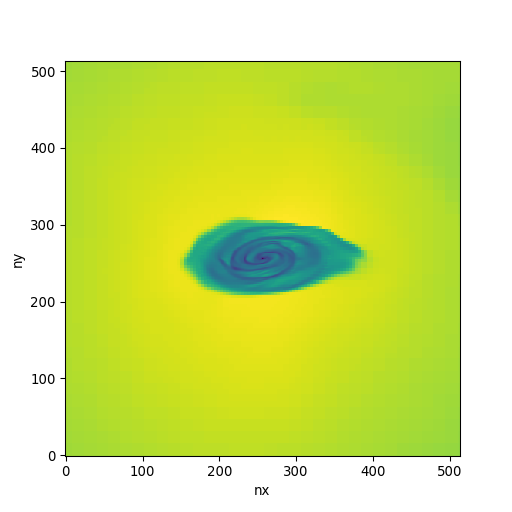}
\includegraphics[scale=0.34]{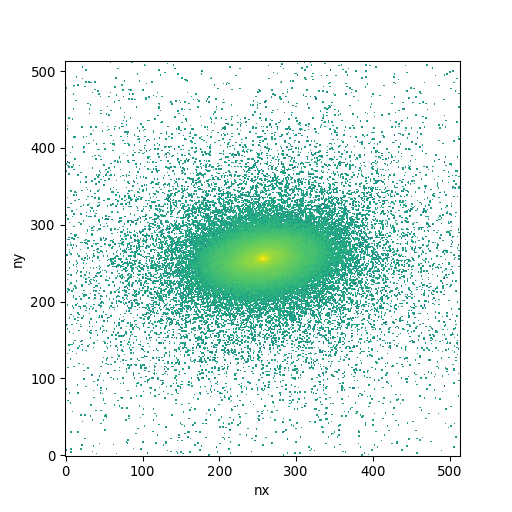}
\caption{Left panel: projected density map of of a zoom-in cosmological simulation of a Milky-Way-size halo with radiative cooling and no feedback. Middle panel: projected temperature map. Right panel: stellar surface density map. Credit figure and simulation: R. Teyssier.}
\label{fig:nofbkgal}
\end{figure}

For halos with masses $10^9 < M_{\rm vir} < 10^{12}$~M$_{\odot}$, we expect the gas to cool down quickly right after the virial shock, and prevent the onset of proper thermal equilibrium. Without pressure support, the gas contracts even more, and reaches much smaller scales closer to the halo center. Be reassured, we don't expect to form a black hole in the center, as another key player will now make an entrance: angular momentum. The gas infall is indeed not perfectly spherically symmetric. Some small asymmetries imprinted by large scale tidal torques induce a net angular momentum so that each collapsing spherical shell is very weakly rotating. As collapse proceeds owing to the strong cooling, angular momentum is conserved and the initially very small tangential velocity increases steadily. At some point, the centrifugal force associated to this increasing tangential velocity will be strong enough to balance gravity. The collapsing gas reaches the so-called centrifugal barrier and collapse stops. The final tangential velocity in the emerging disk can be computed via the centrifugal equilibrium equation:
\begin{equation}
\frac{v_\theta^2}{r} = \frac{G M_{\rm tot}}{r^2}
\end{equation}
where $M_{\rm tot}(r)$ is the cumulative mass corresponding to the NFW model. This gives naturally:
\begin{equation}
v_\theta^2 = \frac{G M_{\rm tot}}{r} \simeq V_{\rm circ}^2
\end{equation}
Cooling breaks thermal equilibrium and promotes the formation of disks in centrifugal equilibrium, for halos in the mass range $[10^9,10^{12}]$~M$_\odot$ that we know as galaxies \cite{Mo.1998}. For larger halos, cooling is less efficient, so we expect less massive galaxies. We show in Figure~\ref{fig:nofbkrun} the projected gas density, gas temperature and the stellar surface density of a simulated zoom-in halo including dark. matter and gas dynamics and also radiative cooling. We see dense cold gas clumps in different location in the halo, with a massive central clump particularly dense and cold. We show in Figure~\ref{fig:nofbkgal} a zoomed image of the central 50~kpc of the halo where a thin rotating gaseous and stellar disk appears. We can see prominent spiral arms. Is this galaxy realistic? To answer this question, we need to model star formation.

\subsection{Simple Model for Star Formation}

We have now reached the point where our galaxy formation theory stops being a self-consistent computational approach where everything can be modeled {\it a priori}: dark matter dynamics solving Vlasov-Poisson equations, gas dynamics solving Euler-Poisson equations and radiative cooling using quantum mechanics and atomic physics. We now need to make luminous stars out of the dense gas that accumulates in our rotating disks. Then can we really bridge the gap with cosmological observations and predict the properties of central and satellite galaxies. The problem is that star formation theory is also an unsolved problem in astrophysics. Many believe that galaxy formation and star formation are tightly interconnected and can only be truly understood via the same complete theory. So how do we connect our simulations with observations of galaxies without a proper theory of star formation? We adopt, as often, a pragmatic, empirical approach.

Observations of nearby galaxies indicate that disk galaxies follows an empirical relation between the star formation rate density and the gas density, known as the Kennicutt relation \cite{Kennicutt.1998}:
\begin{equation}
\Sigma_{\rm SFR}=(2.5\pm0.7)\times 10^{-4} \left( \frac{\Sigma_{\rm gas}}{{\rm M}_{\odot}{\rm pc}^{-2}}\right)^{1,4} [{\rm M}_\odot{\rm yr}^{-1}{\rm kpc}^{-2}]~~~{\rm if}~~~\Sigma_{\rm gas}>1~{\rm M}_{\odot}{\rm pc}^{-2},
\end{equation}
and zero otherwise. This observed relations has suggested simulators to adopt for their star formation recipe a so-called Schmidt law \cite{Schmidt.1959}:
\begin{equation}
\dot{\rho}_*=\epsilon_{\rm ff}\frac{\rho}{t_{\rm ff}}
~~~{\rm if}~~~\rho > \rho_*.
\end{equation}
The local gas free-fall is used to compute the time scale for a gas clump to collapse under its own weight, and the 2 free parameters $\epsilon_{\rm ff}$ and $\rho_*$ are calibrated so that the simulation reproduces the observed Kennicutt relation. Current simulations with resolution between 10 and 100~pc require usually $\rho_*$ between 0.1 and 10H/cc and $\epsilon_{\rm ff} \simeq 0.01$ \cite{Naab.2013}. Adopting this pragmatic method indeed produces spiral galaxies such as the one in Figure~\ref{fig:nofbkgal}, with nice rotating disk and a star formation rate consistent with nearby observations. The problem is: these galaxies are not realistic in the sense that the stellar mass is way too high. In the example shown here, the stellar mass is higher than $10^{11}$~M$_\odot$, with a stellar mass to halo mass ratio around 0.15, close to the universal baryon fraction $f_b=16\%$ and quite far from what abundance matching requires, which is around $3.5\%$ \cite{Agertz.2011,Kretschmer.2019}. The key missing pieces of the galaxy formation puzzle, as suggested by the 2 slopes in the abundance matching formula of \cite{Behroozi.2013}, are supernovae feedback at small halo masses and SMBH-driven feedback (also known as Active Galactic Nuclei) at large halo masses. This will be two topics of the next Chapter.

%% file: chapter4.tex
\section{Subgrid Models for Galaxy Formation}

\subsection{Introduction}

In the previous Chapter, we have identified one of the key limitations of present day galaxy formation model, namely our inability to model star formation from first principles. Abundance matching offers one extreme solution, trying to directly connect empirically the total stellar mass in the galaxy to the host halo mass, using the observed stellar mass function and the theoretical halo mass function. The Schmidt law offers another alternative, trying to calibrate empirically the local star formation rate to the observed Kennicutt relation of nearby galaxies, connecting their gas surface density to their local star formation rate surface density. Both models are based on empirical laws and therefore lack predictive power. 

The key difficulty in developing a predictive model is the fact that we don't resolve the physical scales required to model the collapse of molecular clouds (or the molecular cores they contain) into individual stars. Another obvious difficulty we have to face is the fact that we don't have a complete theory of star formation yet. We will nevertheless discuss in this Chapter some attempts to model star formation, in the context of interstellar turbulence being the main driver of star formation. 

As explained in the previous Chapter, another important ingredient to model galaxy formation is the formation and evolution of Supermassive Black Holes (SMBH) at the center of galaxies. This idea was introduced to explain the quenching of star formation in massive galaxies after major mergers driven star bursts. It is also supported by multiple observations of jets from bright central galaxies hosting active galactic nuclei at the center of large galaxy clusters. Most of these so-called predictive models are in fact not predictive at all, as they depend on crucial and poorly understood parameters. This is why these parameters need to be calibrated on several key observables. 

\subsection{Mean-Field Equations for Turbulence}

We have briefly discussed turbulence in the previous Chapter as the physical process allowing us to describe velocity fluctuations in galactic halos, deviating from strict hydrostatic equilibrium. These fluctuations can be seen as random, following some unknown probability density function. They average to zero, so that the flow is indeed static in an average sense, but they provide some turbulent pressure support, hence modifying the flow on scales larger than the fluctuation scales. We present now a powerful mathematical formalism to include turbulence in our model and how they modify the original equations into a new set of equations called mean-field equations. These new equations are sometimes called Reynolds-Averaged Navier-Stokes (RANS) equations \cite{Reynolds.1895,LANDAU.1987,Pringle.2007}. 

A key difference between our derivation and what is done traditionally in science and engineering applications of turbulence is that we are dealing here with highly compressible flows. We need to account explicitly for density and pressure fluctuations, on top of the more standard velocity fluctuations. We define the mean flow variables as $\overline{\rho}$, $\overline{\bm v}$ and $\overline{P}$, where the averaging process corresponds to a spatial scale $\ell$ and a temporal scale $\tau$ much smaller than the macroscopic scales defined for example by the pressure scale height $H_P$ but also much larger than the microscopic scales so that the Euler equations are still valid on these intermediate scale. We called this intermediate scale quite naturally the {\it mesoscopic scale}, with the following condition:
\begin{equation}
\lambda_{\rm coll} \ll \ell \ll H_P
\end{equation} 
In the context of numerical simulations, we can define unresolved scales as $\ell \le \Delta x$, so that the natural scale at which one averages unresolved fluctuations is the grid resolution $\Delta x$. Subgrid models are precisely trying to model the effect of these unresolved scales on resolved, macroscopic scales. We can now define the density fluctuations $\rho^\prime$ and the pressure fluctuations $P^\prime$ by:
\begin{equation}
\rho = \overline{\rho} + \rho^\prime~~~{\rm and}~~~P = \overline{P} + P^\prime
\end{equation}
We however define the velocity fluctuations differently, using the so-called {\it Favre average}:
\begin{equation}
\overline{\bm v} = \frac{\overline{\rho {\bm v}}}{\overline{\rho}}~~~{\rm and}~~~{\bm v} = \overline{\bm v} + {\bm v}^\prime
\end{equation}
This is the mass-weighted mean of the velocity, or the volume-weighted mean of the momentum divided by the volume-weighted mean of the density. Let's derive now the mean-field mass conservation equation. We start with the mass conservation equation for  unresolved scales which reads:
\begin{equation}
\frac{\partial \rho}{\partial t} + {\bm \nabla} \cdot \left( \rho {\bm v} \right) = 0
\end{equation}
We then average the equation itself, exploiting the fact that now time and space derivative are taken on scales much larger than the fluctuations scales, and therefore commute with the averaging operator. We get immediately:
\begin{equation}
\frac{\partial \overline{\rho}}{\partial t} + {\bm \nabla} \cdot \left( \overline{\rho}~ \overline{{\bm v}} \right) = 0
\end{equation}
This is the mean-field continuity equation. Let's try now to derive the mean-field momentum equation. We start with the momentum equation at unresolved scales;
\begin{equation}
\frac{\partial}{\partial t}\left( \rho {\bm v}\right) + {\bm \nabla} \cdot \left( \rho {\bm v} \otimes {\bm v} + P\mathbb{I}\right) = \rho {\bm g}
\end{equation}
The problematic term in the momentum flux can be simplified using the definition of the velocity fluctuations:
\begin{equation}
\rho {\bm v} \otimes {\bm v} = \rho \left( \overline{\bm v} + {\bm v}^\prime \right) \otimes \left( \overline{\bm v} + {\bm v}^\prime \right)
\end{equation}
Taking the average, we get:
\begin{equation}
\overline{ \rho {\bm v} \otimes {\bm v}} = \overline{\rho} ~\overline{\bm v} \otimes \overline{\bm v} 
+ \overline{ \rho {\bm v}^\prime \otimes {\bm v}^\prime }
\end{equation}
where we used the relations:
\begin{equation}
\overline{\rho \overline{\bm v}} = \overline{\rho}~  \overline{\bm v}~~~{\rm and}~~~\overline{\rho {\bm v}^\prime}=0
\end{equation}
We obtain the mean-field momentum equation:
\begin{equation}
\frac{\partial}{\partial t}\left(\overline{ \rho}~ \overline{\bm v}\right) + {\bm \nabla} \cdot \left( \overline{\rho}~ \overline{\bm v} \otimes \overline{\bm v} + {\mathbb P}_{T} + \overline{P}\mathbb{I}\right) = \overline{\rho} {\bm g}
\end{equation}
where we introduced the turbulent pressure tensor:
\begin{equation}
{\mathbb P}_{T} = \overline{ \rho {\bm v}^\prime \otimes {\bm v}^\prime }
\end{equation}
We see that the mean-field equation for momentum conservation is different from the corresponding original Euler equation by precisely the presence of this new term. In order to compute the pressure tensor, we need to know the statistics of the velocity fluctuations, which is hard because these are defined at unresolved scales. We can however evaluate the magnitude of these fluctuations deriving a new equation describing the time evolution of a new variable called the kinetic energy density of turbulence defined by:
 \begin{equation}
K_{T} = \frac{1}{2} \overline{ {\rho v^\prime}^2 }
 \end{equation}
Using the same methodology, a more involved calculation (left to the reader as an exercise) leads to the turbulent kinetic energy equation:
\begin{equation}
\frac{\partial }{\partial t}\left( K_T\right) +{\bm \nabla}\cdot \left( K_T \overline{\bm v} \right) + {\mathbb P}_T : {\mathbb G} + {\bm \nabla } \cdot {\bm Q}_T = - \epsilon_T
\end{equation}
where we introduce the turbulent dissipation term $\epsilon_T$ and the turbulent heat flux vector ${\bm Q}_T$:
\begin{equation}
{\bm Q}_{T} = \frac{1}{2}\overline{ \rho {v^\prime}^2{\bm v}^\prime }
\end{equation}
The resemblance to the non-LTE fluid equations is striking. We can interpret the effect of these unresolved velocity fluctuations as similar to the effect of particle collisions far from LTE conditions. In the traditional picture of a turbulent flow with small vortices moving in random directions, it is tempting to model these new terms using this analogy between mesoscopic vortices and microscopic particles. This is the main motivation of the Boussinesq approximation, also known as {\it mixing length theory}, where we model the turbulent pressure tensor and the heat flux as:
\begin{equation}
{\mathbb P}_T = P_T{\mathbb I} - \overline{\rho} \nu_T \left( {\mathbb G} + {\mathbb G}^T -\frac{2}{3}\left({\bm \nabla} \cdot \overline{\bm v}\right){\mathbb I} \right)
~~~{\rm and}~~~
{\bm Q}_T = - \overline{\rho} \nu_T {\bm \nabla}\left( \frac{K_T}{\overline{\rho}}\right)
\end{equation}
where $\nu_T$ is the turbulent diffusion coefficient. The isotropic turbulent pressure is defined as $P_T = \overline{\rho} \sigma^2_T = \frac{2}{3}K_T$, where we introduced the 1D turbulent velocity dispersion $\sigma_T$. The mixing length theory models the diffusion and the dissipation as:
\begin{equation}
\nu_T = \ell \sigma_T~~~{\rm and}~~~\epsilon_T = \frac{\sigma_T}{\ell} K_T
\end{equation}
In stellar interiors, mixing length theory for convective flows adopts for the mixing length $\ell \simeq H_P$ \cite{Bohm-Vitense.1958}. In the context of numerical simulations of galaxy formation, a natural choice is $\ell \simeq \Delta x$ \cite{Schmidt.2011}. This is obviously a free parameter of the model. More sophisticated subgrid models for turbulence are still being developed in science and engineering, and are usually tailored for particular applications, with free parameters (such as $\ell$ here) being adjusted to experiments or observations. This formalism, applied to numerical simulations is called Large Eddies Simulations (LES) where we model explicitly only the large, resolved eddies (or vortices), the small eddies being modeled at the subgrid level using the additional terms in the mean-field equations \cite{SMAGORINSKY.1963}.

Interestingly, the pressure tensor in the mean-field momentum equation has a turbulent diffusion coefficient $\nu_T = \ell \sigma_T$. We know from the previous Chapter that the numerical diffusion coefficient is $\nu_{\rm num} \simeq \Delta x \left| v \right|$, which is usually of the same order or even larger. This demonstrates that the additional terms in the mean-field equations are not necessary and can be ignored. This lazy approach is called Implicit LES (ILES), while the previous approach is called Explicit LES (ELES) \cite{Margolin.2006}. In case of ILES, the attentive reader would ask: Why bother then? The turbulent subgrid model we just introduced is actually a crucial ingredient in designing predictive models of star formation.

\subsection{Gravo-Turbulent Star Formation Models}

As explained in a previous Chapter, star formation is described empirically by the Kennicutt relation, which relates the SFR surface density and gas surface density by $\Sigma_{\rm SFR} \propto \Sigma_{\rm gas}^{1.5}$. The Kennicutt relation seems to hold at scales as small as $1~{\rm kpc}$, supporting the idea that star formation can be modeled locally as a volumetric Schmidt law:
\begin{equation}
	\dot{\rho}_* = \epsilon_{\rm ff} \frac{\rho}{\tau_{\rm ff}} \quad {\rm with} \quad \tau_{\rm ff} = \sqrt{\frac{3\pi}{32 G \rho}}
	\label{eq:schmidtlaw}
\end{equation}
where $\dot{\rho}_*$ is the local SFR density and $\epsilon_{\rm ff}$ is the SFE per local freefall time $\tau_{\rm ff}$. This model naturally explains the power law index of $1.5$ in the empirical Kennicutt relation. Equation~\ref{eq:schmidtlaw} is often combined with a fixed density threshold to obtain a simple star formation recipe which can be used in cosmological simulations. The parameter $\epsilon_{\rm ff}$ is chosen to match the observed Kennicutt relation in nearby resolved galaxies, typically resulting in a value $\simeq 1-2\%$. However, the SFE in the local Universe cannot necessarily be extrapolated to earlier epochs. This motivates us to develop a predictive star formation recipe which does not rely on empirical calibration. One approach is to consider the properties of the density fluctuations at unresolved scales $\ell \le \Delta x \simeq 10-100~{\rm pc}$. We extrapolate the turbulence spectrum to unresolved scales assuming Burgers turbulence:
\begin{equation}
	\sigma(\ell) = \sigma_{\rm T} \left( \frac{\ell}{\Delta x} \right)^{1/2}
	\label{eq:burger}
\end{equation}
where $\ell$ is the spatial scale. This scaling law differs from the well-known Kolmogorov law, $\sigma \propto \ell^{1/3}$, which applies for subsonic turbulence. Burgers turbulence is supported by observations in the local interstellar medium, known as Larson's relation. The turbulence transitions from supersonic to subsonic at the sonic scale $\ell_{\rm s}$ where the turbulent velocity dispersion (Eq.~\ref{eq:burger}) is equal to the sound speed. Defining the turbulent Mach number $\mathcal{M}_{\rm T} = \sigma_{\rm T}/c_{\rm s}$, the sonic scale is $\ell_s = \Delta x / \mathcal{M}_{\rm T}^2$. Below the sonic scale, density fluctuations are weak, and a gas cloud can be treated as a quasi-homogeneous region.  We assume that each parcel of gas which is gravitationally unstable at the sonic scale will eventually collapse and form stars. On intermediate scales between the sonic scale and the resolution scale, density fluctuations can be significant. As demonstrated by numerical simulations, the density PDF in a supersonic turbulent medium is well-described by a log-normal distribution \cite{Federrath.2012}:
\begin{equation}
	p_V(s) = \frac{1}{\sqrt{2\pi \sigma_s^2}} \exp ( -\frac{\left(s - \overline{s}\right)^2}{2 \sigma_s^2} )
	\label{eq:mffpdf}
\end{equation}
Subscript $V$ means the function gives the volume fraction of gas at that density. It is normalized using the condition:
\begin{equation}
	\int_{-\infty}^\infty p_V(s) \dd s = 1 \quad {\rm and} \quad \int_{-\infty}^\infty \rho p_V(s) \dd s = \overline{\rho}
\end{equation}
where $s = \ln(\rho / \overline{\rho})$ is the logarithmic density, $\sigma_s$ is its standard deviation, and $\overline{s} = -1/2\sigma_s^2$ is its mean. Numerical simulations also suggest a simple analytic form to $\sigma_s$ using non-magnetized, isothermal turbulence simulations forced by an Ornstein-Uhlenbeck process \cite{Padoan.2011}:
\begin{equation}
	\sigma_s^2 = \ln ( 1 + b_{\rm T}^2 \mathcal{M}_{\rm T}^2 )
	\label{eq:sigs}
\end{equation}
The parameter $b_{\rm T}$ describes the turbulence forcing in the simulations. For purely solenoidal (divergence-free) forcing, $b_{\rm T} = 1/3$. For purely compressive (curl-free) forcing, $b_{\rm T}=1$. In most simulations, the forcing parameter is set to a constant value. However, we can also derive the forcing parameter using the local mean-field velocity field, defining the following expression for ratio of power in compressive to solenoidal forcing modes:
\begin{equation}
	\psi = \frac{P_{\rm comp}}{P_{\rm sol}} = \frac{({\bm \nabla \cdot \overline{\bm v}})^2}{({\bm \nabla} \times \overline{\bm v})^2}
	\label{eq:power}
\end{equation}
and a simple analytic form to $b_{\rm T}$ inspired by detailed turbulence simulations \cite{Federrath.2010a2}: 
\begin{equation}
	b_{\rm T} \simeq \frac{1}{3} + \frac{2}{3} \left( \frac{\psi}{\psi + 1} \right)^3
	\label{eq:bturb}
\end{equation}
Assuming that star-forming cores are homogeneous spheres with diameter $\ell_{\rm s}$, the gravitational stability condition is $\alpha_{\rm vir, core} \ge 1$, where $\alpha_{\rm vir, core}$ is the virial parameter of the core given by:
\begin{equation}
	\alpha_{\rm vir, core} = \frac{2 E_{\rm kin}}{-E_{\rm grav}} = 
	\frac{15}{\pi} \frac{c_{\rm s}^2 + \sigma(\ell_{\rm s})^2}{G \rho \ell_{\rm s}^2}
	\label{eq:alphavir_first}
\end{equation}
The gravitational stability condition can alternatively be expressed as a condition on the density $\rho \le \rho_{\rm crit}$, where
\begin{equation}
	\rho_{\rm crit} = \frac{15}{\pi} \frac{2 c_{\rm s}^2 \mathcal{M}_{\rm T}^4}{G \Delta x} = \alpha_{\rm vir} \overline{\rho} \frac{ 2 \mathcal{M}_{\rm T}^4 }{1 + \mathcal{M}_{\rm T}^2}
\end{equation}
and where $\alpha_{\rm vir}$ is the virial parameter of the cell given by:
\begin{equation}
	\alpha_{\rm vir} = \frac{15}{\pi} \frac{c_{\rm s}^2 + \sigma_{\rm T}^2}{G \overline{\rho} \Delta x^2} = \frac{15}{\pi} \frac{c_{\rm s}^2}{G \overline{\rho} \Delta x^2} (1 + \mathcal{M}_{\rm T}^2)
	\label{eq:alphavir}
\end{equation}
The corresponding critical logarithmic density writes:
\begin{equation}
	s_{\rm crit} = \ln \left[ \alpha_{\rm vir} \frac{2 \mathcal{M}_{\rm T}^4}{1 + \mathcal{M}_{\rm T}^2} \right]
\end{equation}
It depends on the cell size via the cell virial parameter $\alpha_{\rm vir}$. At higher resolutions, a larger density is required to become gravitationally unstable, but gas will naturally reach those higher densities as it collapses down to the smaller cell size. The model breaks down when the entire cell is subsonic $\mathcal{M}_{\rm T} \le 1$ and the sonic scale is larger than the resolution scale. In this case, the density PDF should be interpreted as the probability distribution for the density across the entire cell and the gravitational stability condition (Eq.~\ref{eq:alphavir_first}) should use the resolution scale rather than the sonic scale. We smoothly interpolate between both stability conditions by defining a modified critical logarithmic density:
\begin{equation}
	s_{\rm crit} = \ln \left[ \alpha_{\rm vir} \left( 1 + \frac{2 \mathcal{M}_{\rm T}^4}{1 + \mathcal{M}_{\rm T}^2} \right) \right]
	\label{eq:scrit}
\end{equation}
If each unstable gas parcel collapses in one freefall time and converts all its mass into stars, then the local SFR is given by an explicit and now predictive formula:
\begin{equation}
	\dot{\rho}_* = \int_{s_{\rm crit}}^\infty \frac{\rho}{\tau_{\rm ff}(\rho)} p(s) \dd s = \epsilon_{\rm ff} \frac{\overline{\rho}}{\tau_{\rm ff}(\overline{\rho})}
\end{equation}
where $\epsilon_{\rm ff}$ is the local SFE per free-fall time given by \cite{Hennebelle.2011x6j}:
\begin{equation}\begin{split}
	\epsilon_{\rm ff} =\ & \int_{s_{\rm crit}}^\infty \frac{\tau_{\rm ff}(\overline{\rho})}{\tau_{\rm ff}(\rho)} \frac{\rho}{\overline{\rho}} p(s) \dd s\\
	=\ & \frac{1}{2} \exp( \frac{3}{8} \sigma_s^2 ) \left[ 1 + {\rm erf} \left( \frac{\sigma_s^2 - s_{\rm crit}}{\sqrt{2 \sigma_s^2}} \right) \right]
	\label{eq:mff}
\end{split}\end{equation}
This concludes the derivation of the gravo-turbulent star formation model used in many recent galaxy formation simulations \cite{Kimm.2015s66,Rosdahl.2018,Martin-Alvarez.2018,Trebitsch.2020}. Interpreting this model is no easy task. We will try and summarize its main properties. First, in case turbulence is weak and ${\cal M}_T \ll 1$, the model is completely bimodal, with $\epsilon_{\rm ff} \simeq 0$ if the cell is thermally supported and $\alpha_{\rm vir}>1$ and $\epsilon_{\rm ff} \ge 1$ if the cell is gravitationally unstable with $\alpha_{\rm vir}<1$. With no turbulence, the picture is clear: as a gas fluid element cools down and condense, it starts off with $\alpha_{\rm vir}>1$ and if it becomes cold and dense enough, it will cross the $\alpha_{\rm vir}=1$ threshold and rapidly forms stars within one free-fall time. Second, if turbulence gets stronger, the outcome of the model depends whether the cell is thermally supported or thermally unstable. If the gas is warm and diffuse enough, so that it is thermally stable, turbulence promotes at the subgrid scales some dense clouds that can form stars. So the efficiency is not zero anymore, but higher with $\epsilon_{\rm ff} \simeq 0.01$. We believe this is the regime for large spirals like the Milky Way \cite{Kretschmer.2019}. On the other hand, if the gas is cold and dense, turbulence will reduce the efficiency by providing turbulent support so that the efficiency is not 1 anymore, but lower with $\epsilon_{\rm ff} \simeq 0.1$. We believe this is the regime for starbursts \cite{Kretschmer.2019} or high-redshift galaxies \cite{Andalman.2025}. Interestingly, this gravo-turbulent model allows isolated disk simulations of large spirals to naturally reproduce the Kennicutt relation \cite{Semenov.2016}. More importantly, recent studies have shown that the global star formation efficiency in galaxies is actually not really set by the adopted local efficiency $\epsilon_{\rm ff}$, as long as it is high enough, say $\epsilon_{\rm ff}\ge 0.01$  \cite{Semenov.2017,Semenov.2018}. It is in fact determined by stellar feedback, another important ingredient that we will discuss now.

\subsection{Stellar Feedback}

\begin{figure}
\centering
\includegraphics[scale=0.8]{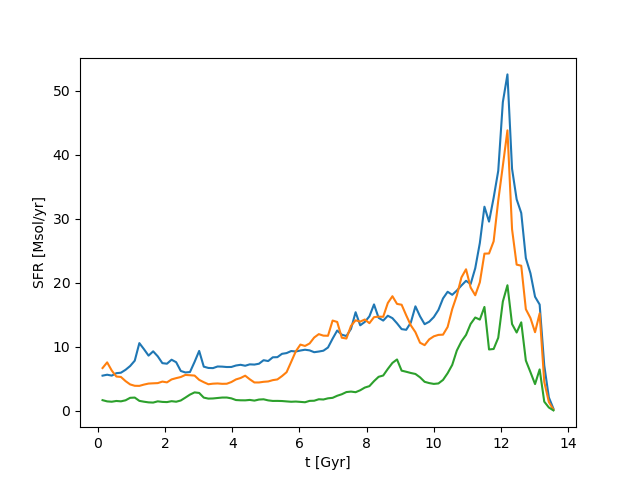}
\caption{Star formation history for 3 different simulations of the same Milky-Way-like galaxy, based on the same dark matter halo simulated in the other Chapters, but varying the supernovae energy from $E_{51}=0$ (blue line), to $E_{51}=2$ (orange line) and finally to $E_{51}=4$ (green line). Credit figure and simulation: R. Teyssier.}
\label{fig:fbksfr}
\end{figure}

In the toolbox of stellar evolution, the most interesting candidates for regulating star formation in the ISM are supernovae explosions \cite{Dekel.1986}. Stars massive enough (larger then 8~M$_\odot$) explodes at the end of their life (between 3~Myr and 20~Myr) in a giant explosion releasing a roughly constant energy of $10^{51}$~erg. The difficulty for cosmological simulators is here again the finite resolution. Typical supernovae remnants are a few parsecs in diameter, requiring here also a subgrid approach. The physics of supernovae blast waves can be summarized in two main phases: the energy-conserving phase and the momentum-conserving phase. The first phase, after the stellar material has been ejected into the ISM and thermalized, can be described by the Sedov similarity solution \cite{SEDOV.1959}. It can be approximated writing a one-zone spherical model where the kinetic energy of the blast wave, identified by its radius $R(t)$, is equal to the initial explosion energy:
\begin{equation}
\frac{4}{3}\pi \rho_0 R^3 \frac{1}{2}\dot{R}^2 = E_0
\end{equation}
where $\rho_0$ is the constant density of the ISM in which the blast wave propagates and $E_0 \simeq 10^{51}$~erg is the explosion energy. 
Solving for this simple ODE gives the simple solution:
\begin{equation}
R(t) \simeq \left(\frac{E_0}{\rho_0}\right)^{1/5} t^{2/5}
\end{equation}
As we have discussed already, gas dynamics is not energy conserving, because of inelastic collisions and radiative cooling. We can estimate the cooling time using the formulae we derived earlier:
\begin{equation}
t_{\rm cool} \simeq \frac{1}{n_0} \left( \frac{T}{10^4~K} \right)^{1/2}~{\rm kyr}
\end{equation}
except we use now a more relevant value for the ISM density $n_0=\rho_0/m_H$ in units of 1~H/cc. For the temperature, we use the Rankine-Hugoniot relations we derive earlier, using here also the strong shock limit with:
\begin{equation}
\frac{k_B T}{m_H} = \frac{3}{5}v_{\rm shock}^2~~~{\rm with}~~~v_{\rm shock} = \dot{R} =\frac{2}{5}\left(\frac{E_0}{\rho_0}\right)^{1/5} t^{-3/5}~~~{\rm or}~~~ T \simeq 6\times 10^8 \frac{E_{51}^{2/5}}{n_0^{2/5}}t_{\rm kyr}^{-6/5}~K
\end{equation}
In our notations, $E_{51}$ is the supernova energy in units of $10^{51}$~erg. We now solve for the cooling radius requiring $t_{\rm cool}(t) = t$, which gives:
\begin{equation}
t_{\rm cool} \simeq 30~{\rm kyr} \frac{E_{51}^{1/8}}{n_0^{3/4}}~~~{\rm and}~~~R_{\rm cool} \simeq 17~{\rm pc} \frac{E_{51}^{1/4}}{n_0^{1/2}}.
\end{equation}
After the blast wave has reached the cooling radius, it enters the so-called snowplow phase, for which the accumulated gas momentum is now conserved. We can write a similar model capitalizing on momentum conservation:
\begin{equation}
\frac{4}{3}\pi \rho_0 R^3 \dot{R} = P_{\rm cool}~~~{\rm with}~~~P_{\rm cool} \simeq 1.4\times 10^5 \frac{E_{51}^{7/8}}{n_0^{1/4}}~{\rm M}_\odot{\rm km/s}~~~{\rm or}~~~R(t) \simeq \left( \frac{3P_{\rm cool}}{\pi \rho_0}\right)^{1/4} t^{1/4},
\end{equation}
where $P_{\rm cool}$ is the momentum of the blast wave at the end of the Sedov phase (hence called the terminal momentum). Note that resolving the energy-conserving phase is crucial to capture the momentum gain of the gas around the exploding star. This requires the spatial resolution to be $\Delta x \ll 10$~pc for moderately dense gas. Traditional supernovae feedback algorithms are releasing a thermal energy dump of $10^{51}$~erg around eligible star particles. If the cooling radius is not properly resolved, thermal energy is radiated away without delivering enough momentum to the gas, therefore failing to achieve the correct effect. The trick here is then to inject directly the terminal momentum into the surrounding gas \cite{Kimm.2015s66,Hopkins.2018}. The momentum-conserving blast wave will propagate until it reaches its final radius for which the ram pressure of the turbulent ISM balances the one of the shock, which means $\dot{R}(t_{\rm final}) \simeq \sigma_T$. 

These equations have been derived assuming a strictly uniform medium. In fact, supernovae explode in a turbulent environment. Detailed numerical simulations of supernovae explosions in turbulent boxes have been performed by \cite{Martizzi.2015}. They showed that the cooling radius depends crucially on the turbulent Mach number ${\cal M}_T$. The exact location of the explosion site is also a crucial parameter: the outcome will be different if the explosion occurs in a dense clump or in a diffuse void, depending also on the exact trajectory of individual stars. Given the complexity of the problem, developing an accurate subgrid model for supernovae explosions is still the topic of intense research. Supernovae explosions have despite everything already played a crucial role in producing much more realistic simulated galaxies. Two key parameters need however to be chosen carefully: how much energy do we release per supernovae and how many stars within a single stellar population are going supernovae? Even though stellar evolution theory has pretty specific predictions for these numbers, we still consider them as free parameters to account for possible uncertainties and finite resolution effect. Hence, here again, galaxy formation simulations need careful calibration of these parameters to fully reproduce observational data.

We show now the results of our same zoom-in simulation of a Milky-Way-like galaxy and show the star formation history of 3 different set of feedback parameters: $E_0 = 0$, $E_0 = 2 \times 10^{51}$ erg and $E _ 0 = 4 \times 10^{51}$ erg with a mass fraction of stars that go supernovae $f_{\rm SN}=0.2$. Only the model with the largest supernovae energy can reproduce the abundance matching results with $M_{*} \simeq 5 \times 10^{10}$~M$_\odot$. Note that this calibration is tightly related to the spatial resolution adopted here namely $\Delta x =100$~pc. A higher resolution simulation would have required a smaller supernovae energy to produce the same final stellar mass.

\subsection{Supermassive Black Holes}

For large galaxies, however, at the center of galaxy groups or galaxy clusters, supernovae feedback cannot regulate star formation anymore. This  is because of the much deeper potential well of the parent halos. Indeed, if one estimate the typical velocity imprinted to gas clouds by supernovae explosions at the end of the energy conserving phase, one get $v_{\rm cool} \simeq 600~{\rm km/s}~~~{\rm where}~~~v_{\rm esc} \simeq \sqrt{c}V_{\rm circ}$. We see that for halos more massive than the Milky Way halo, one has $v_{\rm cool} \le v_{\rm esc}$ so that the gas won't be able to escape the center of the halo very far. Once supernovae feedback fails, the current theory of galaxy formation calls for supermassive black holes. This idea is motivated by two solid observational facts: First, we see multiple evidence of energy release from SMBH hosted by bright central galaxies (BCG) at the center of groups and clusters. Giant X-ray cavities suggest the injection of thermal energy rising buoyantly in the hot intracluster medium (ICM) \cite{Fabian.2002}. Large radio jets seems to emerge from the nuclear region of BCGs and suggest here also the injection of relativistic particles from a central engine \cite{Blandford.2019}. Second, there is a clear correlation between the mass of the central SMBH and the velocity dispersion of the spheroidal component of the host galaxy \cite{Magorrian.1998}. This suggests that the central SMBH and its host galaxy are co-evolving. In addition, there is a population of massive galaxies in which star formation is totally quenched. These red and dead galaxies are difficult to obtain in theoretical models, unless one invokes a central engine disconnected from star formation that can inject energy into the gas and halts star formation, via ejective feedback (ejecting the gas out of the disk) or preventive feedback (preventing the gas from cooling into the disk).

Modeling SMBH and their associated feedback is here again a challenge, as the relevant scales, namely the SMBH horizon $R_G = G M_{\rm BH}/c^2$ or the accretion disk radius (a few times $R_G$) are both very small, around $10^{-3}$~pc for a $M_{\rm BH}=10^{10}$~M$_\odot$. We here again need to develop a subgrid model for SMBH in galaxy formation simulations. These models are articulated around 4 key ingredients \cite{Springel.20057y,McCarthy.2010,Teyssier.2011,Dubois.2012n6c}:
\begin{enumerate}
\item Seeding: Current theory of massive black holes formation rely on only a few poorly understood mechanism. Direct collapse assumes that a giant gas cloud will accumulate and collapse without fragmenting, delivering directly a seed SMBH with mass around $10^4$~M$_\odot$. Dense star clusters at low metallicity provide a favorable environment for stellar remnants of $100$~M$_\odot$ to merge into a few $10^3$~M$_\odot$ SMBH, providing another channel for seed SMBH. It is unclear how to model this from first principles in cosmological simulations so current models just seed directly small early dark matter halos with such seeds, the seed mass being a free parameter.
\item Dynamics: SMBH are usually born in a dense environment, but their subsequent dynamics is unclear. If they are ejected from their parent gas cloud or star cluster, they will have complicated orbits perturbed by other gas cloud. It is unclear if they will manage to stay at the center of the galaxies. The resolution required to properly model all friction mechanisms being enormous, we have to rely on artificial tricks to maintain them at the center, the most brutal one being to maintain the SMBH position at the center of the halo at all time. 
\item Accretion: Once the SMBH are sitting at the center of galaxies, how do we accrete gas onto the accretion disk? Since we do not have the resolution to follow the accretion flow down to sub-parsec scales, we need a subgrid model. The most popular model is the Bondi accretion rate, with an imposed upper bound at the SMBH Eddington rate. Other models have been proposed such as the torque accretion model in galactic disks. None are truly satisfactory.
\item Feedback: Finally, once we know the accretion rate onto the black hole, we release a fraction of the gravitational energy in the form of thermal or kinetic energy. Various models have been implemented in the literature, with multiple free parameters such a the hydrodynamical efficiency, the maximum gas temperature, the jet velocity, etc. All these parameters are poorly motivated theoretically and they are adjusted to reproduce the quenching of massive galaxies in groups and clusters. 
\end{enumerate}
A good review of all these models can be found in \cite{Naab.2013}. Interestingly, when properly calibrated, these models are instrumental in producing realistic galaxies at the high mass end. One of the crucial consequence of such models for cosmology is how they impact the mass distribution on large scales, especially for weak lensing. Even though they are not fully consistent, they are embedded within the framework of galaxy formation simulations, and as such can be used to calibrate baryonification models discussed earlier in the lecture notes, thereby closing the loop.